\documentclass[12pt,preprint]{aastex}

\def\lessim{\mathrel{\hbox{\rlap{\hbox{\lower4pt\hbox{$\sim$}}}\hbox{$<$}}}}
\def\grtsim{\mathrel{\hbox{\rlap{\hbox{\lower4pt\hbox{$\sim$}}}\hbox{$>$}}}}

\shorttitle{M31 Novae}
\shortauthors{Shafter et al.}

\begin{document}

%% LaTeX will automatically break titles if they run longer than
%% one line. However, you may use \\ to force a line break if
%% you desire.

\title{A {\it Spitzer\/} Survey of Novae in M31}

%% Use \author, \affil, and the \and command to format
%% author and affiliation information.
%% Note that \email has replaced the old \authoremail command
%% from AASTeX v4.0. You can use \email to mark an email address
%% anywhere in the paper, not just in the front matter.
%% As in the title, use \\ to force line breaks.

\author{A. W. Shafter\altaffilmark{1}, M. F. Bode\altaffilmark{2}, M. J. Darnley\altaffilmark{2}, K. A. Misselt\altaffilmark{3}, M. Rubin\altaffilmark{1}, and K.~Hornoch\altaffilmark{4}}
\altaffiltext{1}{Department of Astronomy, San Diego State University,
    San Diego, CA 92182}
\altaffiltext{2}{Astrophysics Research Institute, Liverpool John Moores University, Birkenhead CH41 1LD, UK}
\altaffiltext{3}{Steward Observatory, University of Arizona}
\altaffiltext{4}{Astronomical Institute, Academy of Sciences, CZ-251~65~Ond\v{r}ejov, Czech Republic}
%\email{mfb@astro.livjm.ac.uk,mjd@astro.livjm.ac.uk}

%% Notice that each of these authors has alternate affiliations, which
%% are identified by the \altaffilmark after each name.  Specify alternate
%% affiliation information with \altaffiltext, with one command per each
%% affiliation.

%% Mark off your abstract in the ``abstract'' environment. In the manuscript
%% style, abstract will output a Received/Accepted line after the
%% title and affiliation information. No date will appear since the author
%% does not have this information. The dates will be filled in by the
%% editorial office after submission.

\begin{abstract}
We report the results of the first infrared survey of novae in the
nearby spiral galaxy, M31. Both photometric and spectroscopic
observations of a sample of 10 novae
(M31N~2006-09c,
2006-10a,
2006-10b,
2006-11a,
2007-07f,
2007-08a,
2007-08d, 
2007-10a, 
2007-11d, and
2007-11e)
were obtained with the {\it Spitzer\/} Space Telescope.
Eight of the novae were observed with the IRAC
(all but M31N~2007-11d and 2007-11e) and eight with the IRS
(all but 2007-07f and 2007-08a), resulting in
six in common between the two instruments.
The observations, which were obtained between $\sim3$ and $\sim7$
months after discovery,
revealed evidence for dust formation in two of the novae:
M31N~2006-10a and (possibly) 2007-07f, and
[Ne II] $12.8\micron$ line emission in a third (2007-11e).
The {\it Spitzer\/} observations were supplemented with ground-based
optical photometric and spectroscopic data that were used to
determine the speed classes and spectroscopic types of the novae
in our survey. After including data for dust-forming Galactic novae,
we show that
dust formation timescales are correlated with nova speed class in
that dust typically forms earlier in faster novae. We conclude that
our failure
to detect the signature of dust formation in most of our M31 sample
is likely a result of the relatively long delay
between nova eruption and our {\it Spitzer\/} observations. Indeed,
the two
novae for which we found evidence of dust formation were the two
``slowest" novae in our sample. Finally, as expected, we found that
the majority of the novae in our sample belong to the Fe~II spectroscopic
class, with only one clear example of the He/N class (M31N~2006-10b).
Typical of an He/N system, M31N~2006-10b was the fastest nova in our sample,
not detected with the IRS, and just barely detected in three
of the IRAC bands when it was observed $\sim4$ months after eruption.

\end{abstract}


%% From the front matter, we move on to the body of the paper.
%% In the first two sections, notice the use of the natbib \citep
%% and \citet commands to identify citations.  The citations are
%% tied to the reference list via symbolic KEYs. The KEY corresponds
%% to the KEY in the \bibitem in the reference list below. We have
%% chosen the first three characters of the first author's name plus
%% the last two numeral of the year of publication as our KEY for
%% each reference.

%% Authors who wish to have the most important objects in their paper
%% linked in the electronic edition to a data center may do so by tagging
%% their objects with \objectname{} or \object{}.  Each macro takes the
%% object name as its required argument. The optional, square-bracket 
%% argument should be used in cases where the data center identification
%% differs from what is to be printed in the paper.  The text appearing 
%% in curly braces is what will appear in print in the published paper. 
%% If the object name is recognized by the data centers, it will be linked
%% in the electronic edition to the object data available at the data centers  
%%
%% Note that for sources with brackets in their names, e.g. [WEG2004] 14h-090,
%% the brackets must be escaped with backslashes when used in the first
%% square-bracket argument, for instance, \object[\[WEG2004\] 14h-090]{90}).
%%  Otherwise, LaTeX will issue an error. 

\section{Introduction}

Novae are all semi-detached binary systems where a late-type
star transfers mass to a white dwarf companion \citep{war95,war08}.
A thermonuclear runaway eventually ensues
in the material accreted onto the surface of the
white dwarf resulting in the nova explosion \citep{sta08}. The
resulting release of energy ($\sim$~$10^{44}$ -- $10^{45}$ ergs) is
sufficient to expel the accreted envelope and drive substantial mass
loss ($10^{-4}$--$10^{-5}$~M$_{\odot}$) from the system
at high velocities
\citep[$\sim$several hundreds to a few thousand km~s$^{-1}$;][]{bod10}. Novae
exhibit outburst amplitudes of roughly 10 to 20~mag, and can reach
peak luminosities as high as M$_V\sim-10$, second only to GRB and
supernovae in the energetics of their outbursts, but far more frequent
in a given galaxy than either.
Their high luminosities and rates
($\sim50$~yr$^{-1}$ in a galaxy like M31
[Shafter \& Irby 2001; Darnley et al. 2006]),
make novae powerful
probes of the properties of
binary star systems in different (extragalactic) stellar
populations.
The rapid formation of copious amounts of dust in
many novae post-outburst also makes them unique laboratories in which
to explore cosmic dust grain formation
\citep{bod89,eva08,geh08}.

Models show that properties of the eruption (e.g. the peak
luminosity and the decline rate) are strongly dependent on parameters
such as the accretion rate, and the mass and luminosity of the white
dwarf \citep[e.g.,][]{liv92}, some or all of which may vary systematically
with the underlying stellar population.  The strength of the nova
outburst is most sensitive to the mass of the accreting white
dwarf, with the increased surface gravity of a more massive white dwarf
resulting in a higher pressure at the base of the accreted envelope at
the time of thermonuclear runaway and a more violent outburst.
In addition, since a smaller mass of accreted material is
required to achieve the critical temperature and density necessary for
a runaway, nova outbursts produced on massive white dwarfs are
expected to have shorter recurrence times and faster light curve
evolution \citep{hac06}. Further, since the mean white dwarf mass
in a nova system is expected to decrease as a function of the time
elapsed since the formation of the progenitor binary
\cite[e.g.,][]{tut95,pol96}
the proportion of fast and
bright novae is expected to be higher in a younger stellar population,
which should contain on average more massive white dwarfs.

\citet{due90} became the
first to formally postulate the existence of two populations of novae:
a relatively young population of ``disc novae'', which
are found in the solar neighborhood and in the LMC, and ``bulge
novae'', which are concentrated towards the Galactic center and in
the bulge of M31, and are characterized by generally slower outburst
development.
The argument in favor of two nova populations was further
developed by \citet{del92}, who showed that the average
scale height above the Galactic plane for ``fast'' novae ($t_2 < 13$~d
\footnote{$t_2$ is the time in days a nova takes to decline by two
magnitudes from maximum light.})  is smaller than for novae with
slower rates of decline.  At about the same time, \citet{wil92} was
proposing that classical novae could be divided into two classes based
on their spectral properties: specifically, the relative strengths of
either their Fe~II or He and N emission lines. Novae with prominent
Fe~II lines (``Fe~II novae'') usually show P~Cygni absorption
profiles, tend to evolve more slowly, have lower expansion
velocities, and have a lower level of ionization, compared with novae that
exhibit strong lines of He and N (``He/N novae''). In addition,
the latter novae display very strong neon lines, but not the forbidden
lines that are often seen in the Fe~II novae. Following up on
their earlier work, \citet{dvl98} noted that Galactic
novae with well determined distances that were classified as He/N
were concentrated near the Galactic plane, and tended to be faster,
and more luminous compared with their Fe~II counterparts.

Infrared observations offer another avenue for the study of
nova populations.
Models for thermonuclear runaways on the surfaces of white dwarfs
predict that the ejecta will contain not only accreted gas, but
material dredged up from the white dwarf as well. Thus, spectroscopic
analysis of the ejecta offers an opportunity to distinguish between
eruptions that occur on CO white dwarfs and those occurring on more
massive ONe white dwarfs \citep{geh98}.  Eruptions occurring
on CO white dwarfs are expected to produce a significant amount of
dust, mainly in the form of carbon grains that shroud the central
source, and result in warm ($\sim$1000~K) blackbody emission that peaks in
the near IR ($\sim3\micron$) $1-4$ months after eruption
\citep[e.g.,][]{gei70,ney78,bod82,geh95,geh08}.
On the other hand, eruptions seated on the more massive
ONe white dwarfs are believed to produce strong forbidden lines of
neon ([Ne~II] $12.8\micron$ and [Ne~VI] $7.6\micron$), but
little or no dust.  The first observational support for a class of
novae occurring on massive ONe white dwarfs was found more than twenty years
ago when \citet{geh85} discovered that the
[Ne~II] $12.8\micron$ emission line in QU~Vul
($\lambda/\Delta\lambda = 67$) reached a flux level $\sim60$ times
that of the free-free continuum more than four months after
eruption. At the time, this was the strongest $12.8\micron$ emission line
seen in any astrophysical object.  Since that time several additional
Galactic novae with an overabundance of Ne have been discovered, and
they are collectively referred to as ``neon'' novae \citep{geh85}.
It appears that up to one-third of novae
are of this type, with the remainder being CO novae.

Although much can and has been learned from the study of Galactic novae,
the nearby spiral M31, where
more than 800 nova candidates have been discovered over the past century
\citep[e.g.][and references therein]{pie07,sha08},
offers a unique opportunity to study an equidistant sample of novae
from differing stellar environments (bulge and disk) while minimizing
some of the uncertainties that plague Galactic observations (e.g.,
highly variable extinction along the lines of sight to different novae).
With this motivation in mind, we are currently involved in
a multi-wavelength program of photometric and spectroscopic
follow-up observations of novae discovered in M31 \citep{sha10}.
As a part of this effort focusing on the nature of the white dwarfs
(CO vs ONe), we have
obtained both IR photometric and spectroscopic 
observations of a sample of 10 novae in M31 using the
IRAC and IRS instruments on board the {\it Spitzer\/} Space Telescope.
Here, we present the results of our survey.

\section{Observations}

In order to search for dust formation in a sample of M31 novae we
initiated a program of {\it Spitzer\/} Infrared Array Camera
\citep[IRAC;][]{faz04} and Infrared Spectrograph \citep[IRS;][]{hou04}
observations in early 2007 and 2008.
Our observing strategy called for the observations
to be triggered approximately three months after discovery of a nova.
However, due to the
vagaries of {\it Spitzer\/} scheduling, our observations were typically
executed anywhere from 3 to as long as 7 months after eruption.

During the first year of observations, the same 4 novae were observed
with both IRAC and the IRS. The results from the IRS observations were
disappointing, with no line emission detected in any nova, and
continuous emission detected in just two of the novae.
We concluded that most novae had
likely faded beyond detectability by the time
the IRS observations were scheduled. In order to increase the
probability of detection, in the second epoch of observations, we
were able to modify our IRS target list to include
two more recent novae for observation: M31N~2007-11d and 2007-11e,
resulting in line emission being detected in one of these novae.
Our {\it Spitzer\/} observations are summarized in Table~\ref{spitzersummary}.

\subsection{IRAC Observations}

\begin{figure}
\includegraphics[angle=0,scale=.65]{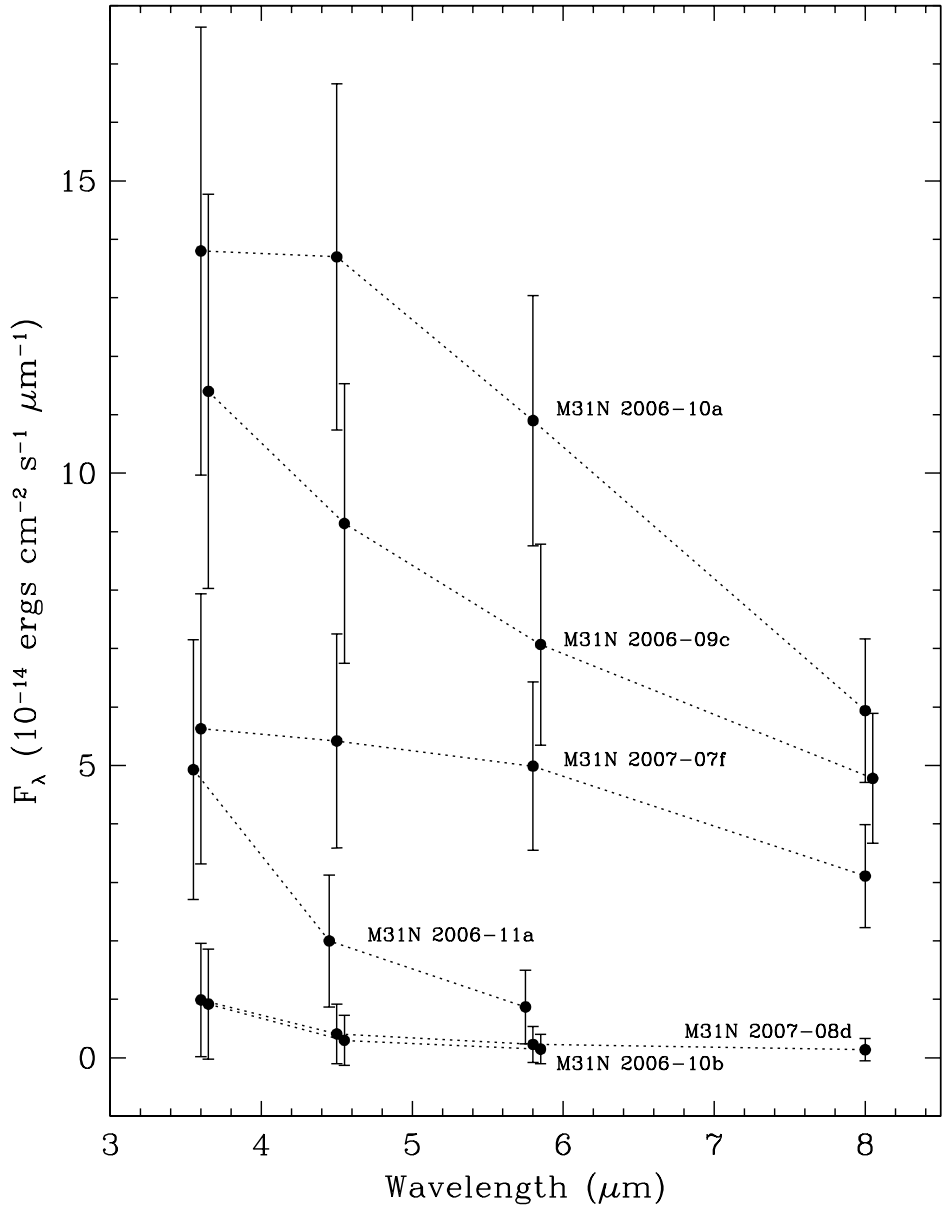}
\caption{The near infrared energy distributions for the six M31 novae
detected by our {\it Spitzer\/} IRAC observations. The data for
M31N~2006-09c, 2006-11a, and 2006-10b have been offset slightly
in wavelength for clarity. Note the apparent IR excess in M31N~2006-10a
and possibly in 2007-07f.
\label{iracsed}}
\end{figure}

IRAC data were obtained in all four IRAC bands ($3.6\micron$, $4.5\micron$, $5.8\micron$, $8.0\micron$).
The
IRAC observations consisted of a 12 point Reuleaux dither pattern with
a medium scale factor. A single 100~sec frame was obtained at each
dither position, resulting in a total of $\sim$1200~sec of exposure in
each band.
Source extraction and photometry were performed on the Post-BCD
(Basic Calibrated Data) mosaics produced by the {\it Spitzer\/} Science Center
(SSC),
pipeline version 18.5.  The extractions and determination of
uncertainties were performed using version 2.1 of the IDL program ATV,
written and maintained by \citet{bar01}.
Due to the crowded nature of the observed fields along with rapidly
varying background levels, it was necessary to use apertures of
minimal size to avoid contamination from neighboring objects.
Therefore we chose an aperture size of radius 2 pixels
for each extraction. We used the nominal positions of the novae
as determined by visual photometric observations, along with an
ATV 3-pixel-radius centering box, to determine the center for each
aperture used in extraction.  Aperture correction factors,
derived by the SSC and provided in the IRAC
Data Handbook 3.0, were then used to estimate the true flux of the
novae based on the measured flux in each aperture.
Finally, we determined the background level for each aperture using
an annulus of inner radius of 2 pixels and outer radius 6 pixels.
The energy distributions for the six novae detected by IRAC are shown
in Figure~\ref{iracsed} and summarized in Table~\ref{iractab}.

\subsection {IRS Observations}

IRS data were obtained in the low resolution SL1 module,
covering a wavelength range of $7.4-14.5\micron$. The IRS data were
obtained using 24 cycles of 60~sec ramps. As each cycle of an IRS
observation consists of two nod positions, this resulted in a total of
$\sim$2880~sec of on-source integration time per target. 

\begin{figure}
\includegraphics[angle=0,scale=.95]{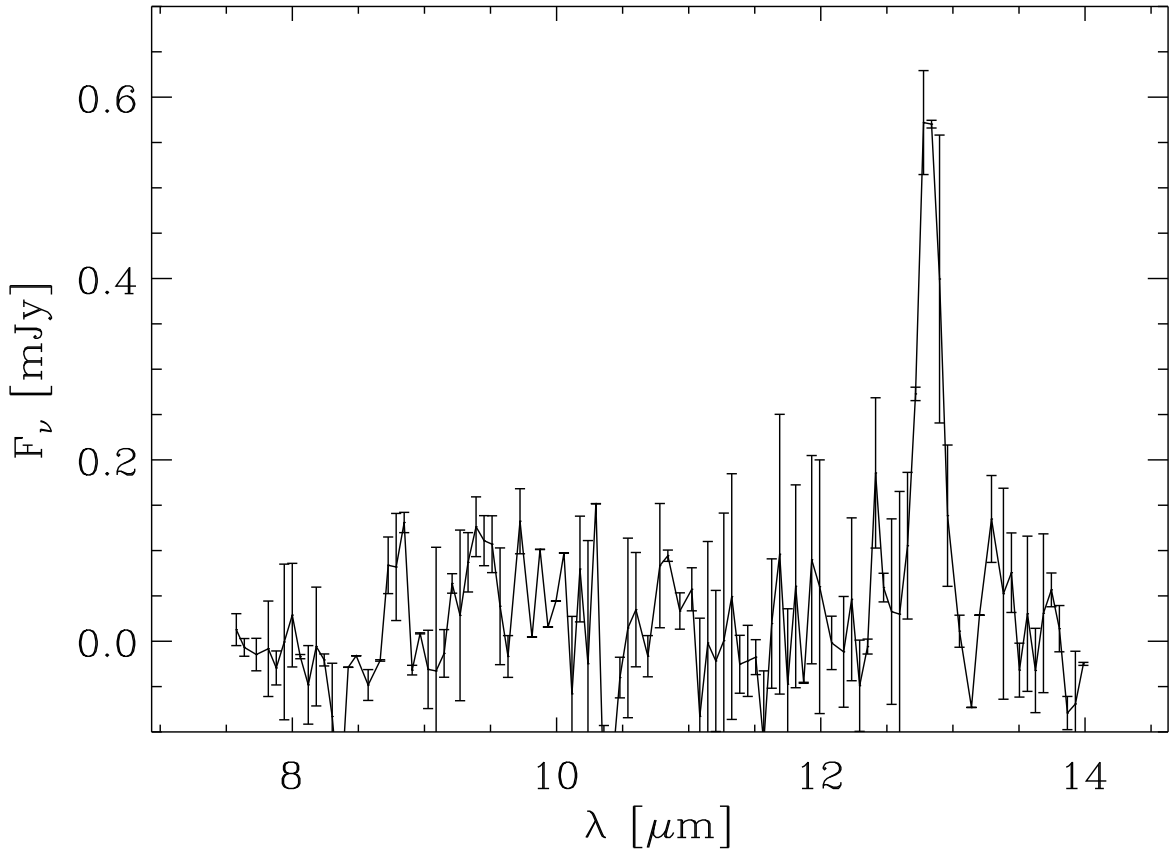}
\caption{The IRS spectrum of M31N~2007-11e showing the [Ne II] $12.8\micron$
line in emission, characteristic of the so-called ``neon novae".
\label{irsed2}}
\end{figure}

The IRS data were analyzed in the SMART package
\citep{hig04} using optimal extraction
\citep{leb10}.
The BCD (Basic Calibrated Data) products were cleaned using the
IRSCLEAN tool with campaign specific masks before all 2D spectra
at a given nod position were combined into a single image.
For our observations,
this resulted in two $\sim$1440~sec cleaned two dimensional
spectra. Prior to spectral extraction, the two nod positions were
subtracted from each other to remove the background and mitigate rogue
pixels not present in the masks.  The spectral extraction was
performed on these background subtracted images.  For each object, we
selected the ``manual optimal extraction'' option in SMART.  The
position of the IRS slit was projected on our IRAC images to verify
that we were extracting spectra for the correct object. A visual
inspection of the two dimensional spectra revealed that only 2 objects
(M31N~2006-09c and 2006-10a) had clear traces at the nominal
source position.  For the remaining objects (with the exception of
M31N~2007-11e, see below), we extracted spectra at the nominal source
position in each nod position.  As expected from the visual inspection
of the two dimensional spectra, no sources were detected. In the case
of M31N~2006-09c, the nova was the only detected source. The spectrum
of M31N~2006-09c was extracted using the optimal extraction algorithm
with a background order of one; even though background subtraction was
accomplished to first order by subtracting nod positions, residual
structure was still present owing to the variation of the local
background between the nod positions. In the case of M31N~2006-10a,
there were two objects detected, the brighter nova
at the nominal position in the slit as well as a fainter source offset
by $\sim$3 pixels ($\sim$5\arcsec). To extract M31N~2006-10a, two
sources were specified and the extraction algorithm was allowed to fit
the trace within one pixel of the nominal position (for the nova) and
between one and four pixels from the nominal position for the
contaminating source. In the case of M31N~2007-11e, no source was
apparent save for a faint line at the position of
[Ne~II] $12.8\micron$.  For M31N~2007-11e, we forced the
extraction aperture to be centered on the nominal source position for
each nod.  For all three objects, spectra extracted at each nod
position were clipped at the order edges and combined.  The spectra
extracted for M31N~2006-09c and 2006-10a are shown in
Figure~\ref{irsed1} and the extracted spectrum of M31N~2007-11e is shown
in Figure~\ref{irsed2}.

\subsection{Optical Photometric Observations}

To complement our {\it Spitzer\/} observations, we attempted
to obtain optical photometric
time-series of several novae in our survey using the
Liverpool Telescope \citep[LT,][]{ste04}, primarily to measure
the peak nova brightness and rate of decline ($t_2$).
The LT data were reduced using a combination of IRAF\footnote{
IRAF is distributed by the National Optical Astronomy Observatory, which is operated by the Association for Research in Astronomy, Inc. under cooperative agreement with the National Science Foundation.} and Starlink software.
These data were calibrated with standard stars from
\citet{lan92} and checked against secondary standards
from \citet{mag92}, \citet{hai94}, and \citet{mas06}.
These observations were augmented by the
extensive photometric database compiled by one of us (KH) as part
of an on-going program to monitor nova light curves in M31.
The latter observations include both survey and targeted
images taken with the 0.35-m telescope of private observatory of KH
at Lelekovice, the 0.65-m telescope of the Ond\v{r}ejov observatory
(operated partly by the Charles University, Prague), and the 0.28-m
telescope of the Zl\'{i}n observatory.
Standard reduction procedures for raw CCD images were applied
(bias and dark-frame subtraction and flat-field correction) using
SIMS\footnote{\tt http://ccd.mii.cz/} and Munipack\footnote
{\tt http://munipack.astronomy.cz/} programs.
Reduced images of the same series were co-added to improve
the S/N ratio (total exposure time varied from ten minutes up to
about one hour).
The gradient of the galaxy background
of co-added images was flattened by a spatial median filter
using SIMS.
These processed images were then used to search for novae.
Photometry was performed using ``Optimal Photometry'' (based on fitting
of PSF profiles) in GAIA\footnote{\tt http://www.starlink.rl.ac.uk/gaia},
and calibrated with standard stars from \citet{lan92}.

\begin{figure}
\includegraphics[angle=0,scale=.65]{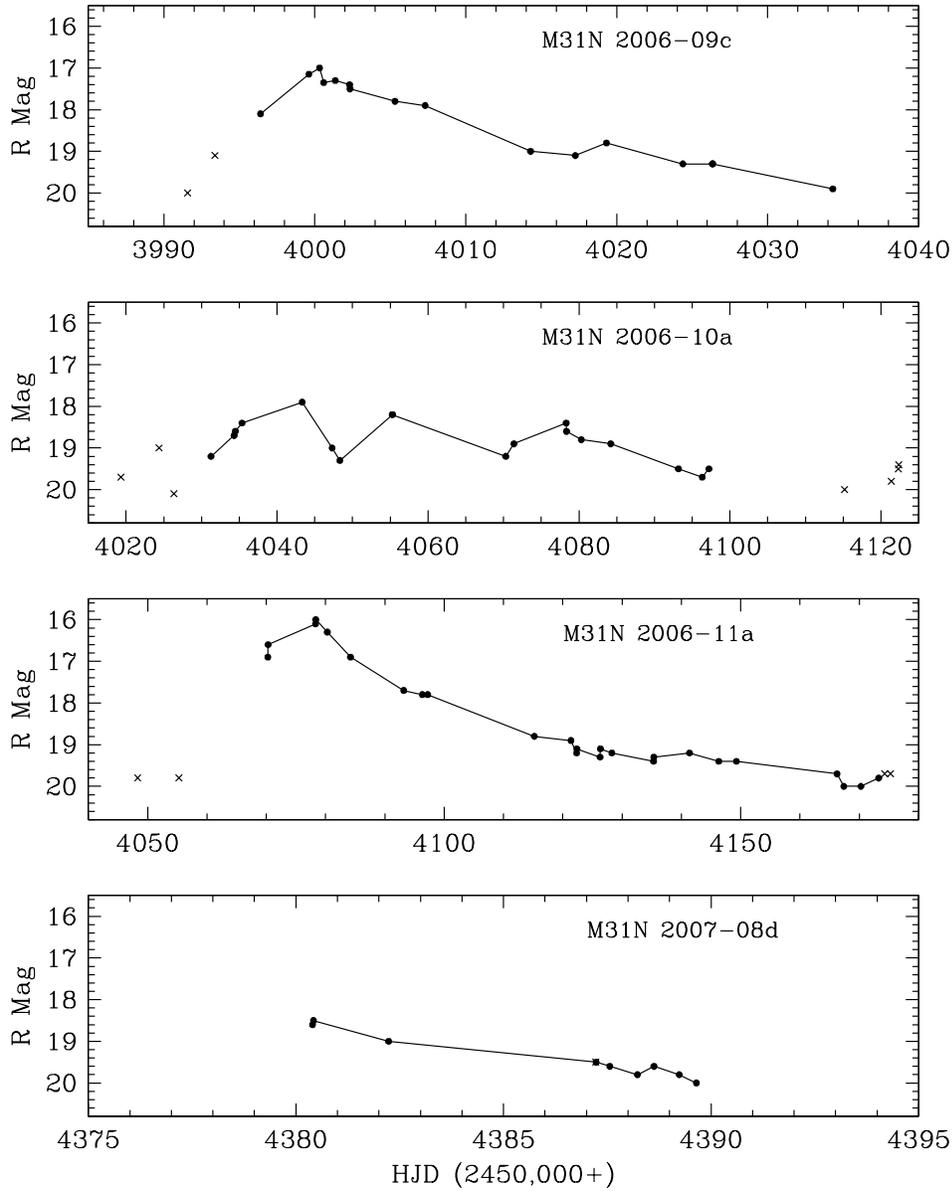}
\caption{$R$-band light curves for four novae observed by {\it Spitzer\/}.
The $\times$ symbols refer to lower limits on the $R$ magnitudes. The light curves yield $t_2(R)=17$~d, $t_2(R)=112$~d, $t_2(R)=21$~d, and $t_2(R)=14$~d for
M31N~2006-09c, 2006-10a, 2006-10b, and 2007-08d, respectively.
\label{lc1}}
\end{figure}

\begin{figure}
\includegraphics[angle=0,scale=.65]{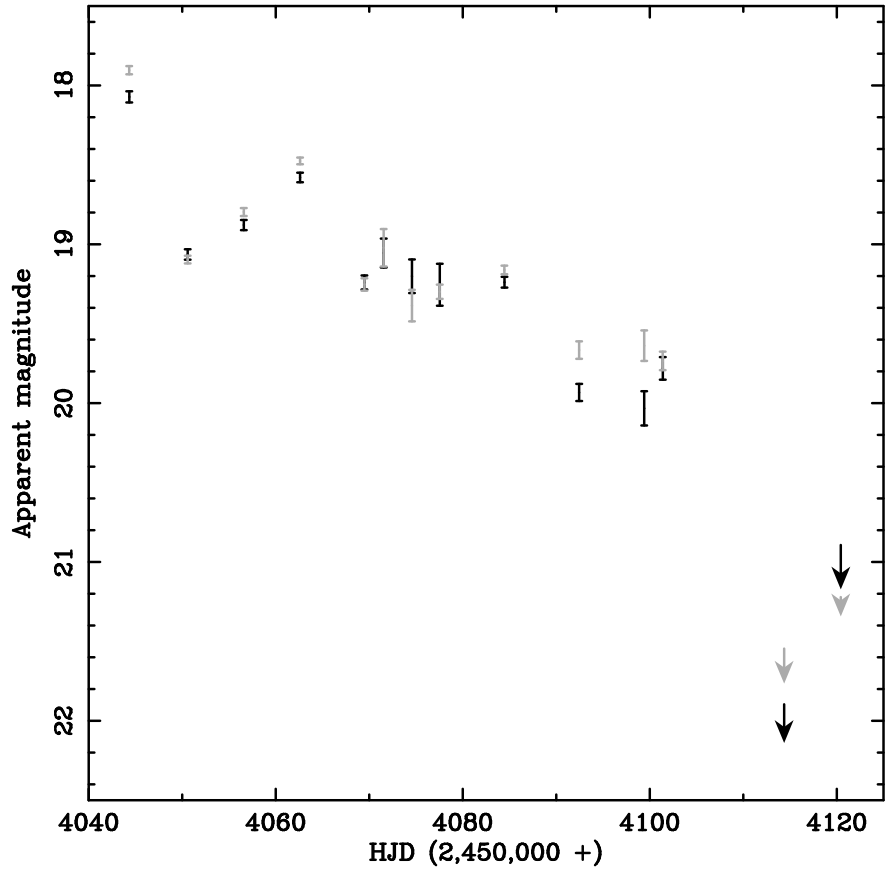}
\caption{The $B$ and $V$ light curves for M31N~2006-10a obtained with the Liverpool Telescope (The $V$ light curve is shown in grey). Arrows are upper limits. From these light curves we derive $t_2(B)=74$~d and $t_2(V)=83$~d.
\label{lc2}}
\end{figure}

\begin{figure}
\includegraphics[angle=0,scale=.65]{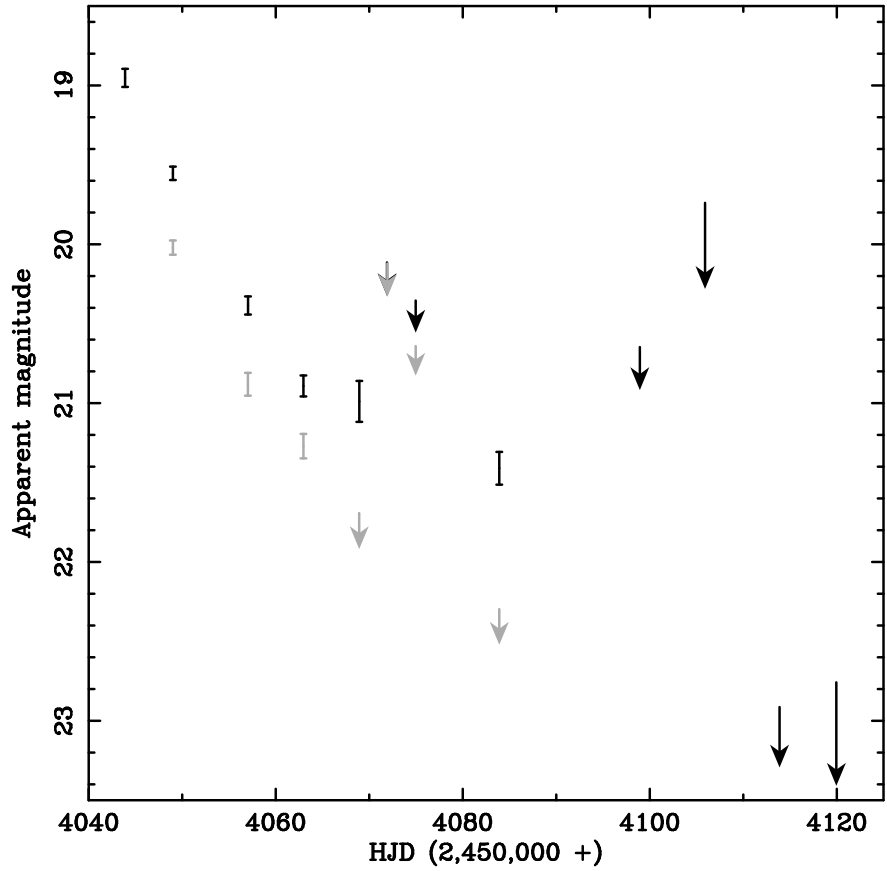}
\caption{The $B$ and $V$ light curves for M31N~2006-10b obtained with the Liverpool Telescope (The $V$ light curve is shown in grey). Arrows are upper limits. Maximum light was missed in $V$, but the $B$ light curve
yields $t_2(B)=20$~d.
\label{lc3}}
\end{figure}

Of the ten novae observed by {\it Spitzer\/} we were successful
in obtaining a series of photometric observations for five of the novae.
The results of these observations are summarized in
Tables~\ref{phot1} -- \ref{phot5}, with the resulting light curves
shown in Figures~\ref{lc1} -- \ref{lc3}.

\subsection{Optical Spectroscopic Observations}

In addition to our ground-based photometric
observations, we also attempted to
obtain optical spectroscopic observations of each
nova using the Low Resolution Spectrograph
\citep[LRS;][]{hil98} on the Hobby-Eberly Telescope (HET).
We used either the $g1$ grating with a
1.0$''$ slit and the GG385 blocking filter covering
4150--11000\,\AA\ at a resolution of $R\sim 600$,
or the $g2$ grating with a 2.0$''$ slit and the GG385 blocking filter,
covering 4275--7250\,\AA\ at a resolution of $R\sim 650$.
When employing the lower-dispersion $g1$ grating, we limited
our analysis to the 4150--9000\,\AA\ wavelength range where the effects of order
overlap are minimal.
The spectra were reduced using standard IRAF routines to flat-field
the data and to optimally extract the spectra.
A summary of all the HET observations is given in Table~\ref{hettab}.
Because the observations were made under a variety of atmospheric
conditions with the stellar image typically overfilling the spectrograph slit,
our data are not spectrophotometric. Thus, the data have been
displayed on a relative flux scale.

\begin{figure}
\includegraphics[angle=0,scale=.65]{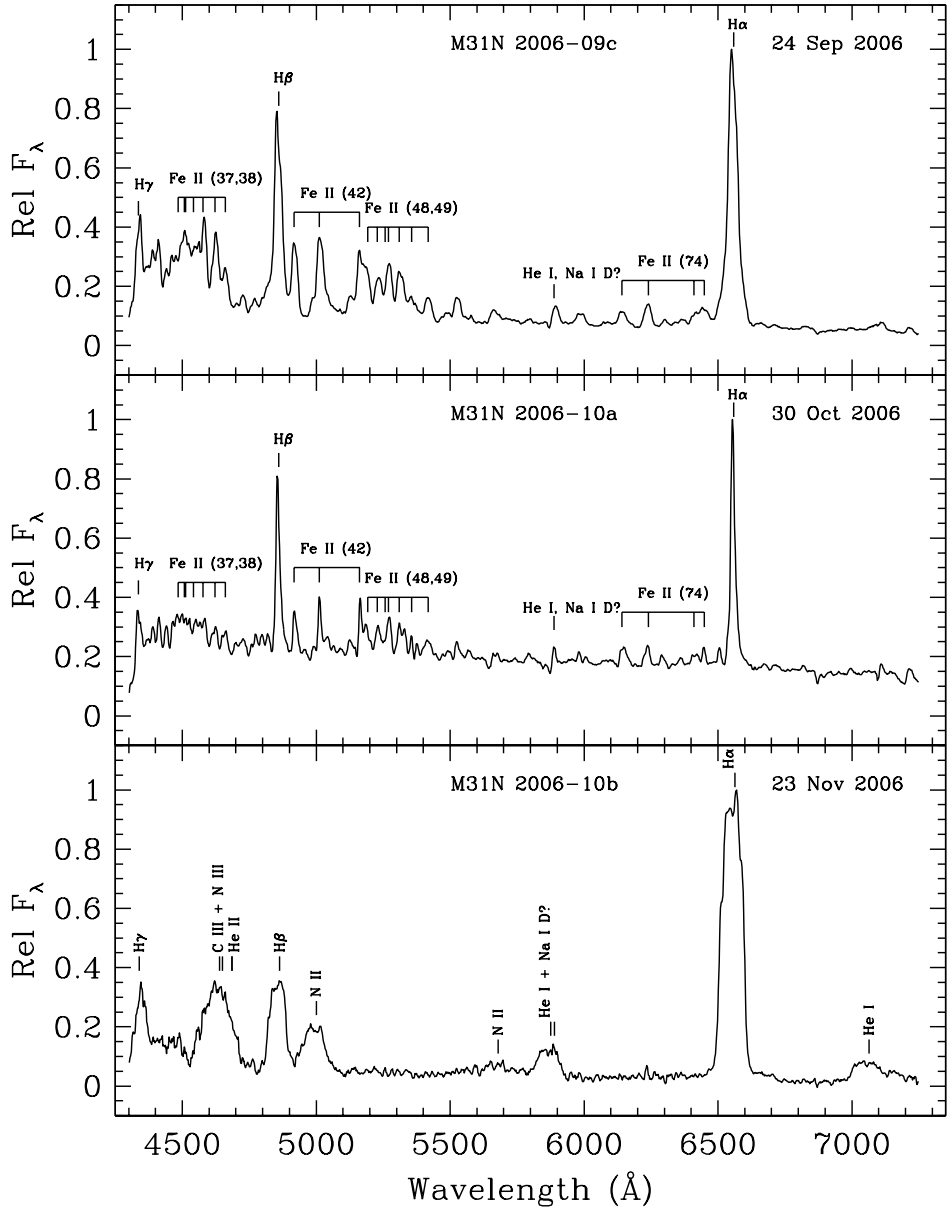}
\caption{HET Spectra of M31N~2006-09c, 2006-10a, and 2006-10b. Note the broad
Balmer, He, and N emission lines in M31N~2006-10b characteristic of the He/N spectroscopic class.
\label{het1}}
\end{figure}

\begin{figure}
\includegraphics[angle=0,scale=.65]{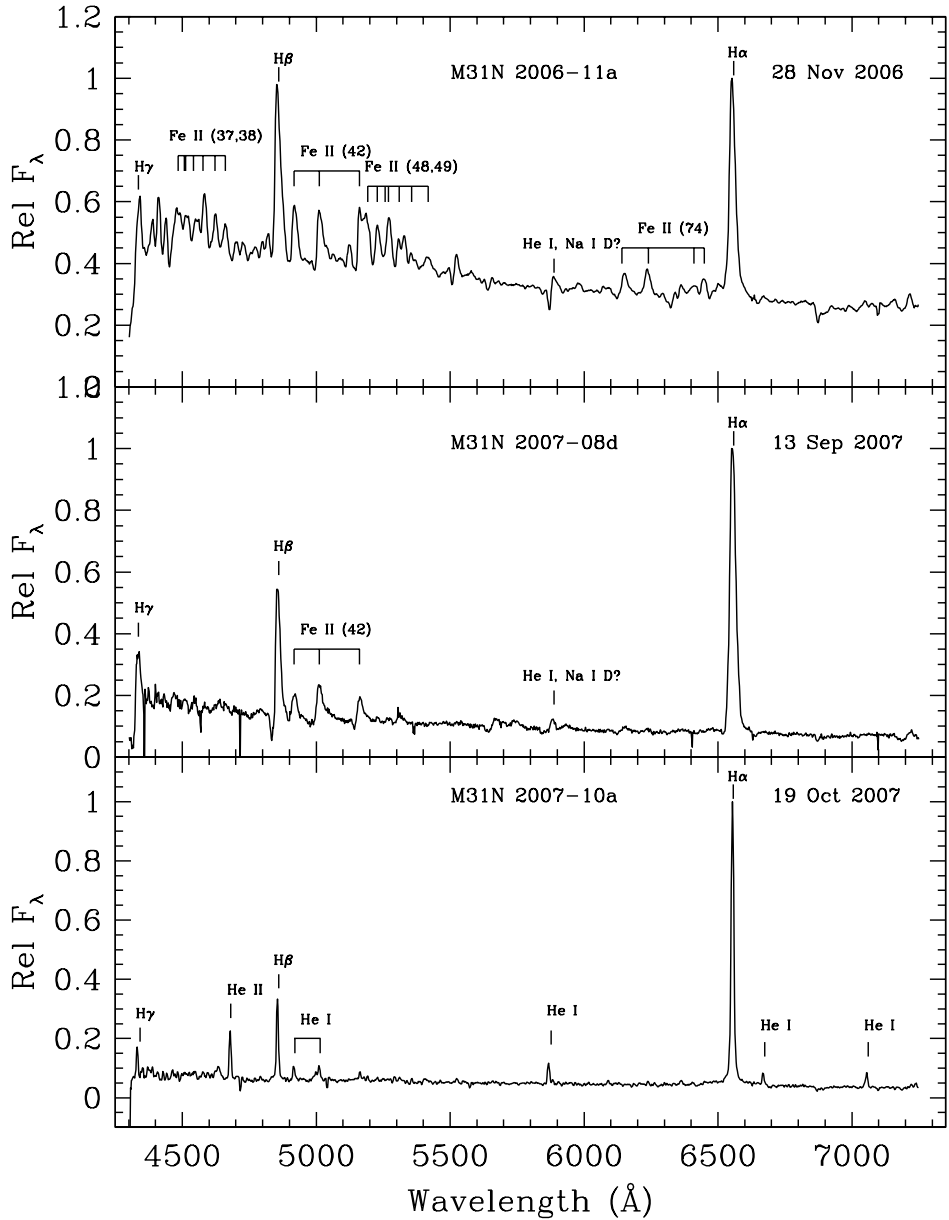}
\caption{HET Spectra of M31N~2006-11a, 2007-08d, and 2007-10a. The spectrum of the latter nova is peculiar in that it has characteristics of both the Fe~II and He/N classes (narrow Balmer and He~I emission lines without significant Fe~II emission).
\label{het2}}
\end{figure}

\begin{figure}
\includegraphics[angle=0,scale=.65]{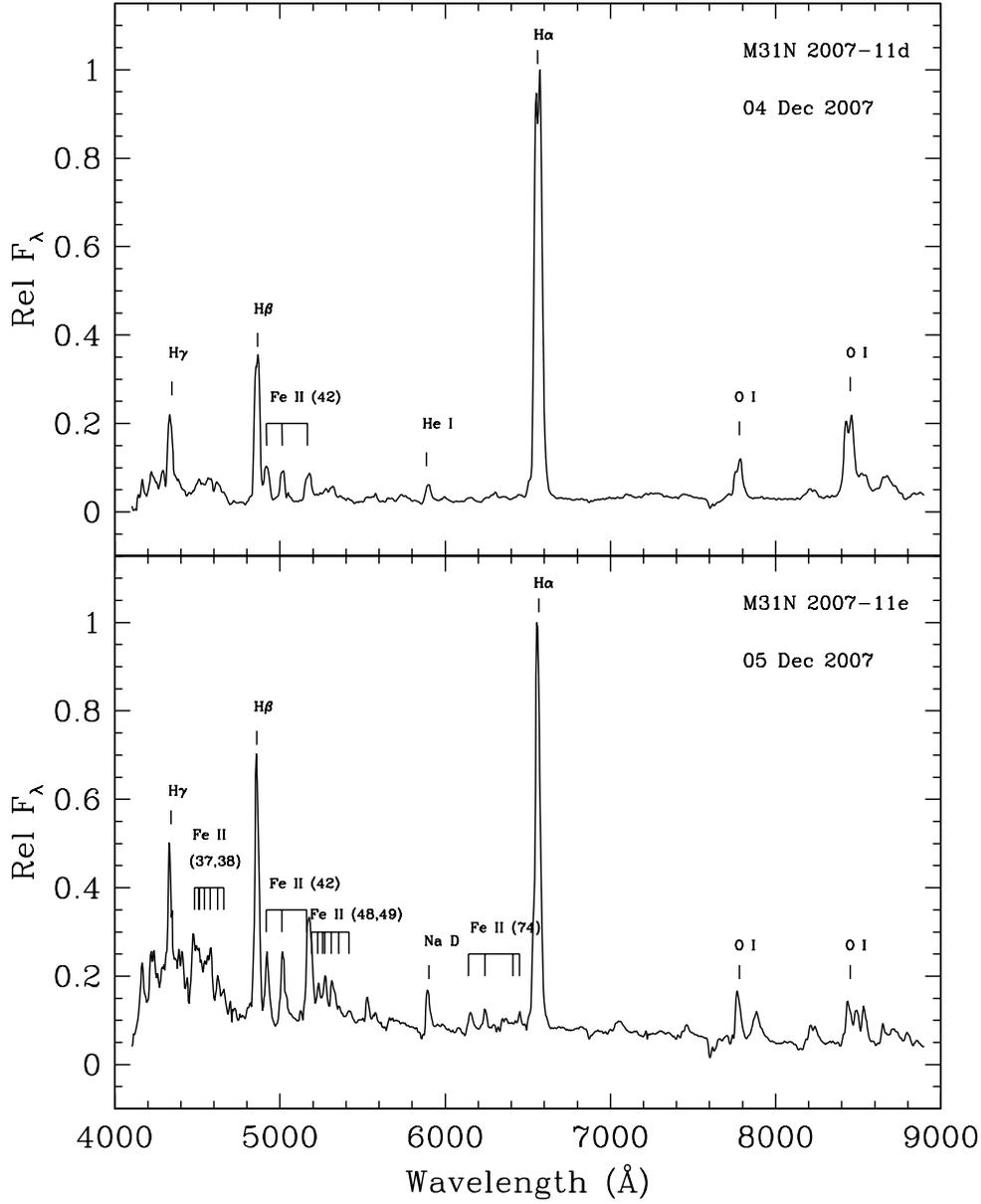}
\caption{HET Spectra of M31N~2007-11d and 2007-11e. Both novae are typical Fe~II systems.
\label{het3}}
\end{figure}

Spectra were successfully
obtained for eight of the ten novae observed by {\it Spitzer\/}.
These data, which were obtained primarily to
ascertain the spectroscopic classes \citep{wil92} and ejection velocities
of the novae,
are shown in Figures~\ref{het1} -- \ref{het3}, and
are summarized in Tables~\ref{emission} and \ref{spatpos}.
Spectroscopic classes for the two novae that we were unable to observe,
M31N~2007-07f and 2007-08a, were
available through spectra obtained
by \citet{qua07} and \citet{bar07}, respectively.

Seven of the ten novae in our sample clearly belong to the Fe~II class,
which is characterized by relatively narrow Balmer and Fe~II emission.
Of the remaining three, M31N~2006-10b was a
``hybrid" nova (a broad-lined Fe~II system that later evolved into
a He/N system), while
M31N~2007-08a was described by \citet{bar07} as an
Fe~II (or possibly a hybrid) nova.
The large Balmer emission line widths seen in M31N~2006-10b are
typical of what is found in He/N novae; however, the
relatively narrow lines exhibited by 2007-08a
are more characteristic of the Fe~II class.
The remaining nova, M31N~2007-10a, is quite peculiar spectroscopically
(see Figure~\ref{het2}). The spectrum displays narrow emission
characteristic of the Fe~II novae, but in addition to Balmer emission,
shows prominent He~I lines rather than Fe~II emission.

\begin{figure}
\includegraphics[angle=-90,scale=.65]{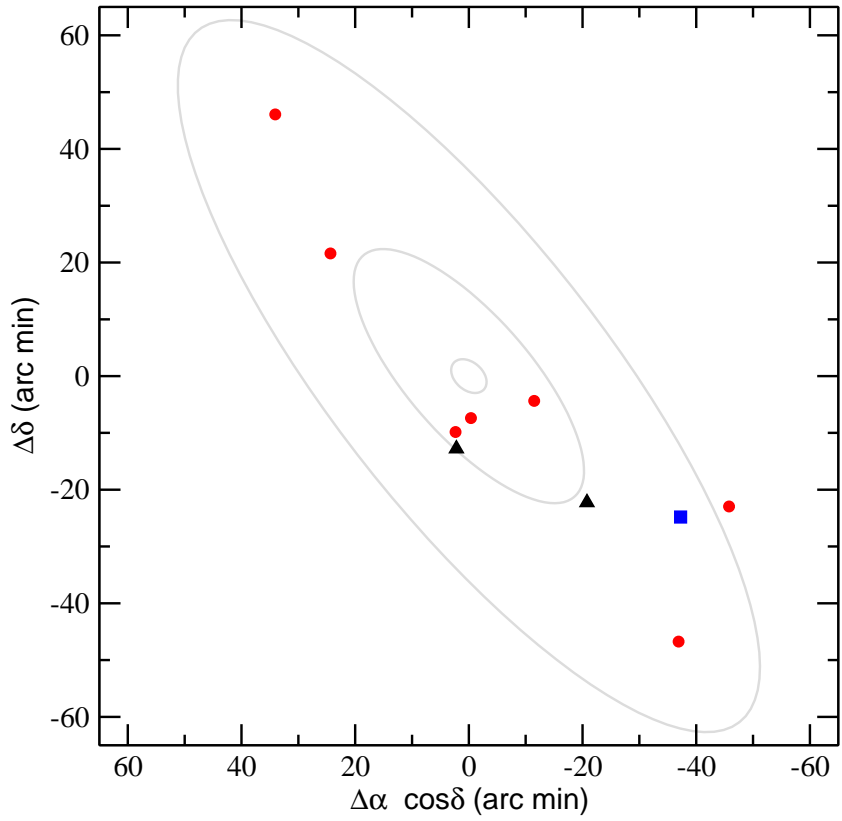}
\caption{The spatial distribution of the 10 M31
novae observed with {\it Spitzer\/}. The Fe~II novae are
indicated by red circles, M31N~2006-10b,
the Hybrid nova that evolved
into a He/N nova is indicated by a blue square, and the
two novae (M31N~2007-08a and 2007-10a)
of uncertain type by black triangles.
\label{spat}}
\end{figure}

The spatial positions of the 10 novae in our {\it Spitzer\/} sample are
shown in Figure~\ref{spat}. As has been demonstrated by \citet{sha07}
from a larger sample of novae,
there is no evidence that the spatial distribution of Fe~II and
He/N novae differ in M31 \citep[see also][]{sha10}.

\section{Discussion}

\begin{figure}
\includegraphics[angle=0,scale=.95]{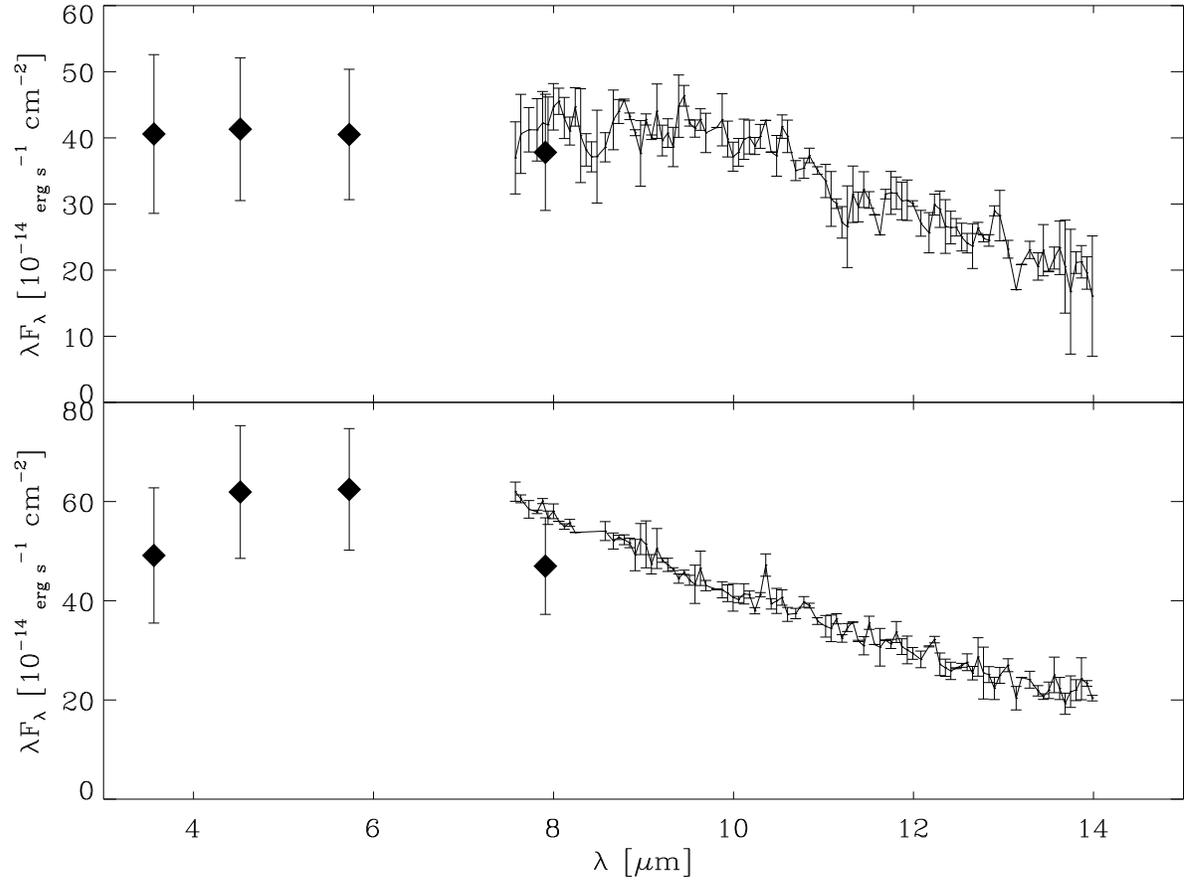}
\caption{The infrared spectral energy distributions for the two
novae detected by both IRAC and the IRS: M31N~2006-09c (upper panel)
and 2006-10a (lower panel). The IR excess, peaking in $\lambda F_{\lambda}$
at $\lambda\sim5\micron$, is obvious in M31N~2006-10a.
\label{irsed1}}
\end{figure}

Of the 8 novae observed by IRAC, only M31N~2006-10a shows
clear evidence of an infrared excess in the IRAC bands,
peaking at $\lambda \sim 4\micron$ (see Fig.~\ref{irsed1}).
M31N~2007-07f shows
evidence for a rather weaker, slightly longer wavelength excess
(see Fig.~\ref{iracsed}). We may associate such excess emission
with that from dust grains condensing in the ejecta of each nova,
as has been observed in many Galactic novae -- see Section 1. 
Both M31N~2006-10a and 2007-07f belong to the Fe~II spectroscopic
class, and both can be considered relatively ``slow" novae
with decay times of $t_2(V)=83$ and $t_2(R)\sim$50 days, respectively.

To further explore the relationship between dust formation timescales
and nova speed class, we have augmented our M31 sample with
available data on
Galactic dust-forming novae.
A summary is presented in Table~\ref{dust}.
In Figure~\ref{tcond} we have plotted the condensation time,
$t_{\rm cond}$, of dust grains in Galactic novae against
$t_{\rm 2}$. Dust condensation times for this plot have been
taken from Table 13.1 of \citet{eva08} and times of dust
breaks in D-type novae from \citet{str10}.
In general, these are consistent within a few days,
but we have taken the earlier quoted time in each case,
if they are both available for a given nova.
Speed class, in terms of $t_{\rm 2}$, has been taken
from \citet{str10}. We have also plotted the two suspected
dusty M31 novae on Figure~\ref{tcond},
where $t_{\rm cond}$ is likely to be an upper limit in
both instances (i.e. we have not necessarily observed the
onset of the dust formation event).
Note that for M31N~2006-10a and 2007-11d,
we have used $t_{\rm 2}$($V$) but as this was unavailable
for the other M31 novae, we have used $t_{\rm 2}$($R$).

There is a clear trend for the Galactic novae in the
sense that the faster novae tend to form grains earlier
than the slow novae. An outlier is nova PW Vul.
\citet{geh88} note that this nova had a very erratic early
optical light curve which resembled DQ Her,
but showed no deep minimum as did the latter,
archetypal dust-forming nova. Indeed PW Vul's dust shell
was also low mass and optically thin and the quoted value
of $t_{\rm cond} = 154$ days was calculated by \citet{geh88}
from the angular expansion rate of the pseudo-photosphere
and the maximum flux of the outburst.
This is consistent with the time of first condensation being
in the observational gap between 100 and 290 days post-outburst.
Also shown as a lower limit on this plot is the dust
condensation time for QU Vul, given by \citet{eva08} as
40--200 days (we will return to this object below).
Marking an extreme point on the plot is nova V445 Pup.
A dust extinction dip is clearly seen in its light curve
and (relatively sparse) infrared photometry thereafter
confirms the presence of an extensive dust shell,
but this is an example of a very rare Helium nova
outburst \citep[see][and references therein]{wou09}.

The apparent correlation between $t_{\rm cond}$ and $t_2$ is perhaps surprising.
If we assume a constant central source luminosity, $L_*$,
and outflow velocity, $v_{\rm ej}$, for a given nova, and that the
grain condensation temperature, $T_{\rm cond}$, is constant from
nova to nova, we can write \citep[e.g.,][]{geh08}:

\begin{equation}
t_{\rm cond} \propto L_*^{0.5}~v_{\rm ej}^{-1}.
\end{equation}

Adopting available empirical relations between $v_{\rm ej}$ and $t_2$, and
between $L_*$ and $t_2$ (the Maximum-Magnitude vs Rate-of-Decline
relation) such as those given in \citet{war08}, we find that
$v_{\rm ej} \propto t_2^{-0.5}$ and $L \propto t_2^{-1}$.
After inserting these relations into eqn~(1),
it appears that $t_{\rm cond}$ should be essentially independent of
nova speed class! Clearly, the empirical correlation shown in Figure~\ref{tcond}
justifies further investigation,
but as pointed out by \citet{eva08},
dust grain nucleation and growth rely on a complex set of parameters.
Finally,
it should be noted that not all novae form detectable amounts of dust,
with examples at both ends of the range of speed
class \citep[e.g. V1500 Cyg versus HR Del;][]{bod89}.
The reasons for this have been explored by various
authors \citep[e.g.,][]{gal77, bod83} but are not fully understood.

Figure~\ref{tIRmax} shows a plot of nova speed class versus
the time of maximum infrared flux, $t_{\rm IRmax}$, as observed in
Galactic novae. Since $t_{\rm IRmax}$ is not available for
our M31 novae, we have instead plotted the speed class
and time after discovery (approximate optical maximum)
for the novae observed with IRAC.
Both M31N~2006-10a and 2007-07f are marked on the plot.
Again there appears to be a trend of increasing $t_{\rm IRmax}$ with
speed class for the Galactic novae, with the exception of QU Vul.
We may note however that this is the time of the maximum flux
observed in the $10\micron$ silicate feature in this
nova by \citet{geh86} at $t = 240$ days,
and not necessarily the time of maximum continuum emission
from grains in the expanding shell, which may have been (much) earlier.
Indeed, an observation on day 206 for this
nova by \citet{smi95} showed that at this time emission from
graphitic dust clearly dominated the $10\micron$ continuum in QU Vul.

From Figure~\ref{tIRmax}, it appears that M31N~2006-10a
may have been observed by {\it Spitzer\/} nearer the time of the
infrared emission maximum than 2007-07f, which again is consistent
with the higher dust shell luminosity and inferred dust
temperature in 2006-10a compared to 2007-07f.
Indeed, M31 2006-10a might have been observed not too
long after $t_{\rm cond}$ at around the time of $t_{\rm IRmax}$.
Evidence for this comes from its position on Figs.~\ref{tcond}
and \ref{tIRmax}, coupled with $T_{\rm BB}$ ($\sim720\pm100$K)
being much lower than $T_{\rm cond}$ ($\sim1200$K).
There is in fact a steep drop in observed effective dust
grain temperature in well observed dusty novae coinciding
approximately with infrared maximum arising from
what \citet{bod83} termed the ``infrared pseudo-photosphere"
in the optically thick (in the infrared) dust shells of prolific
dust formers such as NQ Vul \citep{ney78}.
As the dust formation rate reduces in the expanding shell,
the observed optical depth also reduces and $T_{\rm BB}$
increases again. \citet{mit83} found even better agreement
with observations if grains were subject to some size reduction
around the time of infrared maximum, possibly due to sputtering.
In NQ Vul the temperature minimum lasted from around 62 to 130
days post-outburst. 

All but one of the faster M31 novae lie above the region
occupied by the Galactic novae at infrared maximum
(the  exception being M31N~2006-11a).
The non-detection of dust emission in these objects may then
be due to them either being observed well after dust emission maximum,
or the fact that they did not produce large amounts of dust
in the first place. The latter appears to be the case
for 2006-11a at least, as it does lie in the region of the
plot occupied by the dusty Galactic novae.

We may estimate the total mass of dust in M31N~2006-10a,
$M_{\rm d}$, assuming an isothermal, optically-thin dust shell
of uniform spherical grains of radius $a$ from the equation

\begin{equation}
M_{\rm d} \approx a \rho_{\rm g} d^2 {F_\lambda} \frac{B(\lambda, T_{\rm g})}{Q_{\rm abs}(\lambda, a)},
\end{equation}

\noindent
where $d$ is the distance, $\rho_{\rm g}$ the bulk density of the
grain material, $F_\lambda$ the observed flux density at
wavelength $\lambda$, $B$ the Planck function at $\lambda$ and
grain temperature $T_{\rm g}$, and $Q_{\rm abs}$ the grain
absorption coefficient. If $Q_{\rm abs} \propto \lambda^{-\alpha}$ for
small dielectric absorbers, it can be shown that 

\begin{equation}
T_{\rm g} =\frac{2890}{\lambda_{\rm max}}\Big(\frac{5}{\alpha + 5}\Big),
\end{equation}

\noindent
where $\lambda_{\rm max}$  is the wavelength of maximum emission in $\micron$.

In the case of M31N~2006-10a, the IRS spectrum in Figure\ref{irsed1}
shows no apparent $10\micron$ silicate feature, and the emitting
grains are more likely carbon-based, as seen at some point in the
evolution of most dust-forming novae \citep{eva08,geh08}.
Taking for simplicity graphite spheres of size small compared
with the wavelength of emission, then $\alpha \approx 2$ \citep{eva94}.
Thus with $\lambda_{\rm max} = 4\micron$, $T_{\rm g} = 516$~K,
and if $a = 0.1\micron$,
then $Q_{\rm abs} = 4.72 \times 10^{-2}$
\citep[][although this neglects any temperature dependence
of the absorption coefficient]{dra85}.
Taking $\rho_{\rm g} = 2000$ kg m$^{-3}$ \citep{lov92},
and a distance to M31 of 780 kpc \citep{hol98,sta98}
then $M_{\rm d} \sim 2 \times 10^{-6}$ M$_\odot$.
This is consistent with the range of dust masses derived
for Galactic novae, particularly the more prolific
dust-formers \citep[see e.g.,][]{geh08}.

\section{Summary and Conclusions}

We have conducted the first infrared survey of novae in the nearby spiral, M31,
using the {\it Spitzer\/} Space Telescope.
The primary motivation behind the survey was to determine the feasibility
of using {\it Spitzer\/} observations of M31 novae to study dust formation
property as a function of nova speed class and spectroscopic type.
We observed a total of 10 novae in M31 with {\it Spitzer\/}
over a two year period: 
M31N~2006-09c,
2006-10a,
2006-10b,
2006-11a,
2007-07f,
2007-08a,
2007-08d,
2007-10a,
2007-11d, and
2007-11e.
Eight of these novae were observed with IRAC (all but M31N~2007-11d and
2007-11e) and 8 with the IRS (all but M31N~2007-07f and
2007-08a), with 6 novae observed with both instruments.

Observations of Galactic novae show that dust formation
typically occurs with a timescale, $t_{\rm cond}$,
of between $\sim1$ and $\sim5$ months post-eruption (mean $\sim 2$ months),
depending on the speed class of the nova.
For a typical nova,
the peak infrared signature occurs shortly thereafter, with the time
to infrared maximum, $t_{\rm IRmax}$, averaging about 3 months post-eruption.
Thus, our observing strategy was to schedule our {\it Spitzer\/} observations
approximately 3 months post discovery when the infrared signature due
to dust formation was expected to reach a maximum. Unfortunately, the constraints imposed
by the {\it Spitzer\/} scheduling process did not allow us to time our
observations as precisely as we would have liked, and our observations
occurred anywhere between $\sim3$ and $\sim7$ months post eruption.
Our principal conclusions can be summarized as follows:

\noindent
{\bf (1)} We were able to detect six of the eight novae observed with IRAC.
Of these, only M31N~2006-10a 
showed clear evidence for an infrared excess peaking at 
$\lambda \sim 4\micron$. The IRS spectrum of this nova showed
no evidence of silicate emission features and thus we assume
that the dust was carbon-based in this case.
We were then able to estimate the total mass of dust formed
to be $M_d\sim2\times10^{-6}$~M$_{\odot}$. This is comparable to the
mass of dust found in Galactic novae forming the more extensive dust shells.
Another nova, M31N~2007-07f, showed evidence of possible dust formation
through a weaker infrared excess detected to peak at longer wavelength.
Our observations
of these two novae occurred 116 and 203 days post discovery, respectively.
Thus, it is plausible that
that our observations of M31N~2007-07f occurred after the
peak IR flux was achieved
as suggested by our comparison with the times of infrared maximum in
Galactic dust-forming novae. Both novae are Fe~II systems with relatively
slow light curve evolution: $t_2(V)=82$~d and $t_2(R)\sim50$~d, respectively.

\begin{figure}
\includegraphics[angle=0,width=\textwidth]{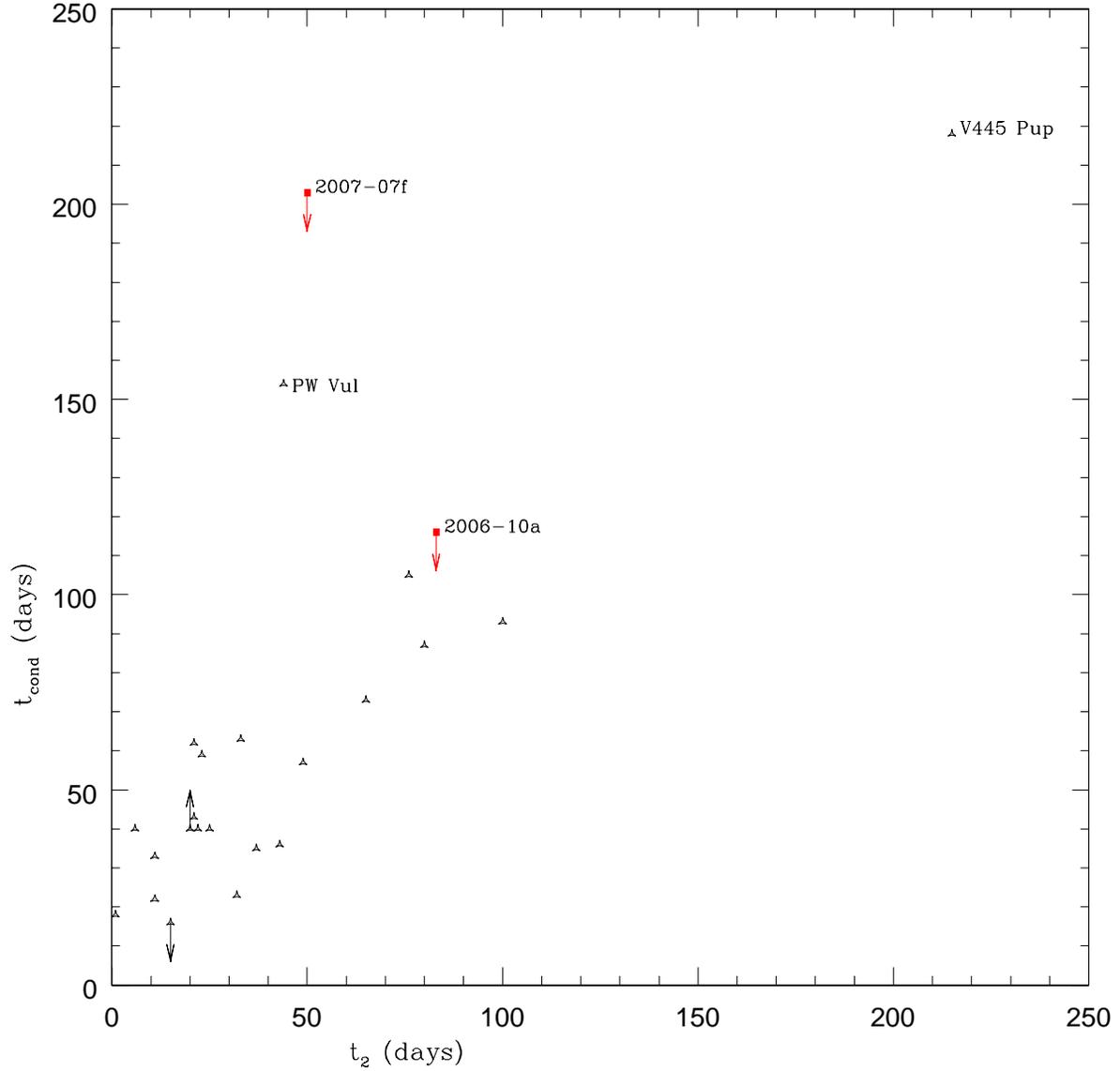}
\caption{Observed condensation time of dust grains in the nova ejecta, $t_{\rm cond}$, versus nova speed class ($t_{\rm 2}[V]$, except for M31N~2007-07f, which is $t_2[R]$). Data for Galactic novae are given as open triangles, with the two suspected dust-forming novae in M31 shown as filled red squares. These, and the special cases of the Galactic novae PW Vul and V445 Pup, are discussed more fully in the text.}\label{tcond}
\end{figure}

\begin{figure}
\includegraphics[angle=0,width=\textwidth]{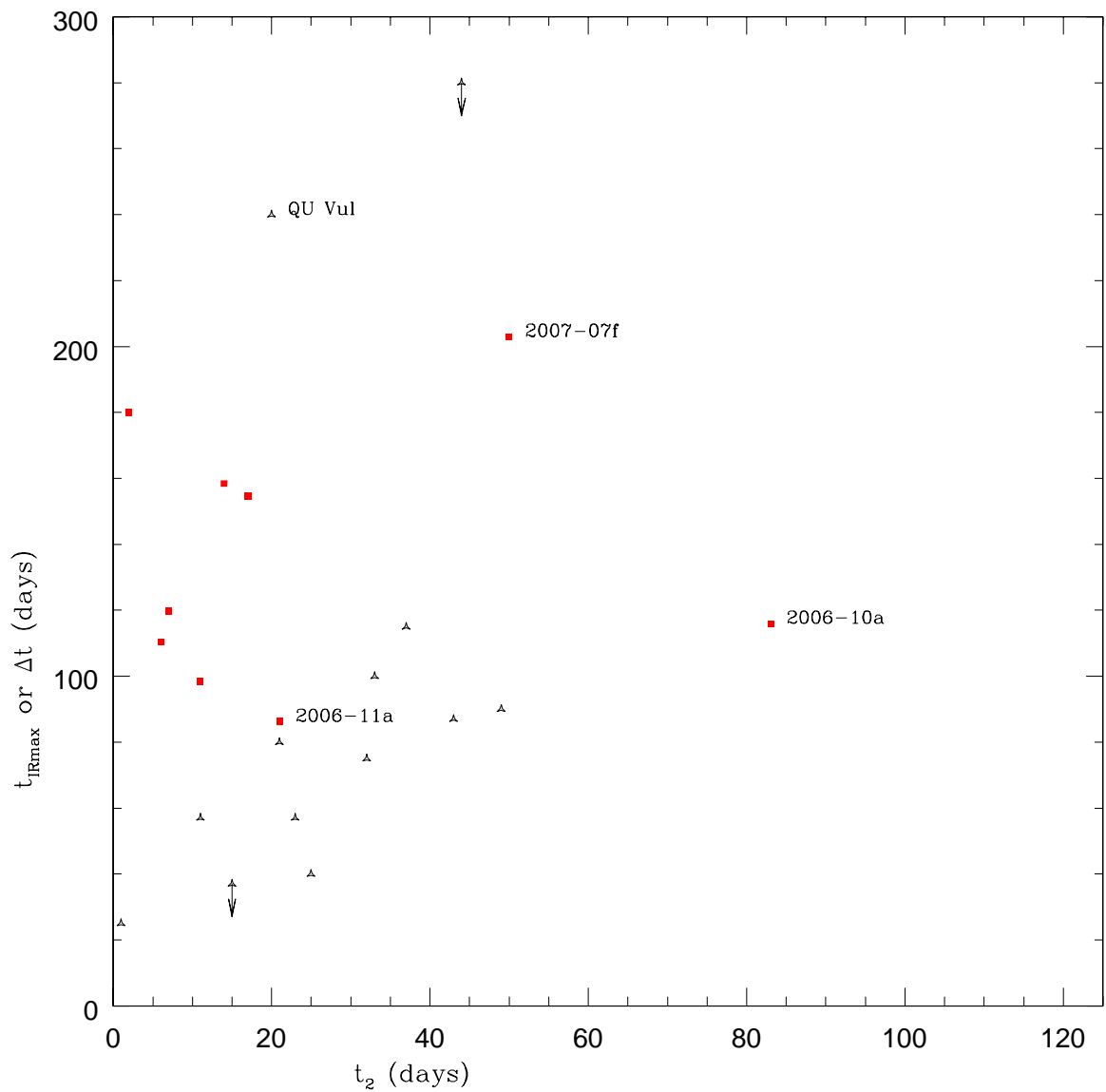}
\caption{Time of infrared maximum, $t_{\rm IRmax}$, versus nova speed class for Galactic novae \citep[open triangles, $t_{\rm IRmax}$ data from][see also Table~\ref{dust}]{eva08}. An outlier is QU Vul, which is discussed more thoroughly in the text. Red squares denote the observations of novae in M31 $\Delta t$ days after discovery (near optical maximum) with objects of particular interest noted (see text for details). Values of $t_2$ are from $V$-band measurements for M31N~2006-10a and the Galactic novae, while the M31 novae are mainly based on $R$-band measurements (again see text for details).}\label{tIRmax}
\end{figure}

\noindent
{\bf (2)} We find a surprising correlation between the condensation time for dust grains ($t_{\rm cond}$)
and nova speed class ($t_2$) for Galactic novae (see Fig.~\ref{tcond}). Although
upper limits, the values of $t_{\rm cond}$ for the
two M31 novae with detected IR excesses are consistent with this correlation.
Most of the M31 novae in our sample, which have $t_2<\sim50$, were likely
observed too long after dust condensation for us to detect an IR excess
in our data. One exception is
M31N~2006-11a, which is a moderate-speed-class, Fe~II nova observed less than 3
months after discovery. A comparison with Galactic dust-forming novae
(see Fig~\ref{tIRmax}) shows that an IR excess likely would have been visible
assuming dust formation had taken place.

\noindent
{\bf (3)} Three of the eight novae observed with the IRS were detected:
M31N~2006~09c and 2006-10a showed only continuum emission, while
2007-11e revealed a [Ne II] $12.8\micron$ emission feature characteristic
of the class of ``Neon Novae". Other than its neon emission, M31N~2007-11e,
is fairly
unremarkable in other respects, being neither a particularly fast or slow
Fe~II system.

Our preliminary survey has demonstrated that IR observations
can be used to detect dust formation in M31 novae.
It is likely that those novae that went undetected
were observed too long
after maximum light when the dust had cooled sufficiently to render
the IR flux below our limit of detection. Future observations,
possibly from the ground,
of a larger sample of novae
will be required to further characterize the relationship between
nova speed class, spectroscopic type, and dust formation timescales.
The correlations we find between speed class and $t_{\rm cond}$ and $t_{\rm IRmax}$
will greatly assist observational planning.

\acknowledgments

The work presented here is
based in part on observations obtained with the
Hobby-Eberly Telescope,
which is operated by McDonald Observatory on behalf of the University
of Texas at Austin, the Pennsylvania State University,
Stanford University, the Ludwig-Maximillians-Universitaet,
Munich, and the George-August-Universitaet, Goettingen.
Public Access time is available on the Hobby - Eberly Telescope
through an agreement with the National Science Foundation.
The Liverpool Telescope is operated on the island of
La Palma by Liverpool John Moores University in the Spanish Observatorio del
Roque de los Muchachos of the Instituto de Astrofisica de Canarias with
financial support from the UK Science and Technology Facilities Council
(STFC). KH is grateful for obtaining and providing of M31
images to P. Ku\v{s}nir\'ak, M. Wolf, P. Zasche, P. Caga\v{s}, and
T. Henych. AWS is grateful to the NSF for support through grant AST-0607682, and
to the University of Victoria for hospitality during his sabbatical leave
while this work was being completed.

%% Use the figure environment and \plotone or \plottwo to include
%% figures and captions in your electronic submission.
%% To embed the sample graphics in
%% the file, uncomment the \plotone, \plottwo, and
%% \includegraphics commands
%%
%% If you need a layout that cannot be achieved with \plotone or
%% \plottwo, you can invoke the graphicx package directly with the
%% \includegraphics command or use \plotfiddle. For more information,
%% please see the tutorial on "Using Electronic Art with AASTeX" in the
%% documentation section at the AASTeX Web site,
%% http://www.journals.uchicago.edu/AAS/AASTeX.
%%
%% The examples below also include sample markup for submission of
%% supplemental electronic materials. As always, be sure to check
%% the instructions to authors for the journal you are submitting to
%% for specific submissions guidelines as they vary from
%% journal to journal.

%% This example uses \plotone to include an EPS file scaled to
%% 80% of its natural size with \epsscale. Its caption
%% has been written to indicate that additional figure parts will be
%% available in the electronic journal.

%% Here we use \plottwo to present two versions of the same figure,
%% one in black and white for print the other in RGB color
%% for online presentation. Note that the caption indicates
%% that a color version of the figure will be available online.
%%

\clearpage

%% If you are not including electonic art with your submission, you may
%% mark up your captions using the \figcaption command. See the
%% User Guide for details.
%%
%% No more than seven \figcaption commands are allowed per page,
%% so if you have more than seven captions, insert a \clearpage
%% after every seventh one.

%% Tables should be submitted one per page, so put a \clearpage before
%% each one.

%% Two options are available to the author for producing tables:  the
%% deluxetable environment provided by the AASTeX package or the LaTeX
%% table environment.  Use of deluxetable is preferred.
%%

%% Three table samples follow, two marked up in the deluxetable environment,
%% one marked up as a LaTeX table.

%% In this first example, note that the \tabletypesize{}
%% command has been used to reduce the font size of the table.
%% We also use the \rotate command to rotate the table to
%% landscape orientation since it is very wide even at the
%% reduced font size.
%%
%% Note also that the \label command needs to be placed
%% inside the \tablecaption.

%% This table also includes a table comment indicating that the full
%% version will be available in machine-readable format in the electronic
%% edition.

\clearpage

\begin{deluxetable}{lccccrl}
\tabletypesize{\scriptsize}
\tablenum{1}
\tablewidth{0pt}
\tablecolumns{7}
\tablecaption{Summary of {\it Spitzer\/} Observations\label{spitzersummary}}
\tablehead{\colhead{} & \colhead{} & \colhead{} & \colhead{Discovery} & \colhead{\it Spitzer\/} & \colhead{} & \colhead{} \\ \colhead{} & \colhead{R.A.} & \colhead{Decl.} & \colhead{$\mathrm{HJD}$} & \colhead{$\mathrm{HJD}$} & \colhead{$\Delta t$} & \colhead{Discovery} \\ \colhead{Nova} & \colhead{(2000.0)} & \colhead{(2000.0)} & \colhead{($2,450,000+$)} & \colhead{($2,450,000+$)} & \colhead{(days)} & \colhead{Mag}}
\startdata
\cutinhead{IRAC}
M31N 2006-09c & 00:42:42.38 & +41:08:45.50 & 3996.64 & 4151.25 & 154.61 & 16.5(w) \\
M31N 2006-10a & 00:41:43.23 & +41:11:45.90 & 4034.31 & 4149.85 & 115.54 & 18.6(R) \\
M31N 2006-10b & 00:39:27.46 & +40:51:10.10 & 4039.59 & 4149.82 & 110.23 & 16.4(w) \\
M31N 2006-11a & 00:42:56.61 & +41:06:18.20 & 4064.97 & 4151.22 &  86.25 & 17.3(w) \\
M31N 2007-07f & 00:38:42.20 & +40:52:56.10 & 4292.48 & 4495.03 & 202.55 & 17.4(w) \\
M31N 2007-08a & 00:40:54.40 & +40:53:50.30 & 4318.80 & 4498.70 & 179.90 & 17.6(w)? \\
M31N 2007-08d & 00:39:30.27 & +40:29:14.20 & 4336.58 & 4495.06 & 158.48 & 18.1(R) \\
M31N 2007-10a & 00:42:55.95 & +41:03:22.00 & 4379.11 & 4498.73 & 119.62 & 16.3(R)? \\
\cutinhead{IRS}
M31N 2006-09c & 00:42:42.38 & +41:08:45.50 & 3996.64 & 4142.04 & 145.40 & 16.5(w) \\
M31N 2006-10a & 00:41:43.23 & +41:11:45.90 & 4034.31 & 4141.99 & 107.68 & 18.6(R) \\
M31N 2006-10b & 00:39:27.46 & +40:51:10.10 & 4039.59 & 4142.08 & 102.49 & 16.4(w) \\
M31N 2006-11a & 00:42:56.61 & +41:06:18.20 & 4064.97 & 4141.95 &  76.98 & 17.3(w) \\
M31N 2007-08d & 00:39:30.27 & +40:29:14.20 & 4336.58 & 4519.34 & 182.76 & 18.1(R) \\
M31N 2007-10a & 00:42:55.95 & +41:03:22.00 & 4379.11 & 4519.30 & 140.19 & 16.3(R)? \\
M31N 2007-11d & 00:44:54.90 & +41:37:39.60  & 4421.02 & 4519.43 &  98.41 & 14.9(w) \\
M31N 2007-11e & 00:45:47.74 & +42:02:03.50  & 4432.58 & 4519.39 &  86.81 & 16.6(w) \\
\enddata
\end{deluxetable}

\begin{deluxetable}{lcccc}
\tabletypesize{\scriptsize}
\tablenum{2}
\tablewidth{0pt}
\tablecolumns{5}
\tablecaption{{\it Spitzer\/} IRAC Photometry\label{iractab}}
\tablehead{\colhead{} & \multicolumn{4}{c}{Flux ($10^{-14}$ ergs~cm$^{-2}$~s$^{-1}$~$\micron^{-1}$)}\\
\colhead{Nova} & \colhead{$3.6\micron$} & \colhead{$4.5\micron$} & \colhead{$5.8\micron$} & \colhead{$8.0\micron$}}
\startdata
M31N 2006-09c & $11.4\pm3.37$ & $9.14\pm2.39$ & $7.07\pm1.72$ & $4.78\pm1.11$\\

M31N 2006-10a & $13.8\pm3.83$ & $13.7\pm2.96$ & $10.9\pm2.14$ & $5.94\pm1.23$\\

M31N 2006-10b & $0.92\pm0.94$ & $0.30\pm0.43$ & $0.15\pm0.25$ & \dots\\

M31N 2006-11a & $4.93\pm2.22$ & $2.00\pm1.13$ & $0.87\pm0.63$ & \dots \\

M31N 2007-07f & $5.63\pm2.31$ & $5.42\pm1.83$ & $4.99\pm1.44$ & $3.11\pm0.88$\\

M31N 2007-08a & \dots & \dots & \dots & \dots\\

M31N 2007-08d & $0.99\pm0.97$ & $0.41\pm0.51$ & $0.23\pm0.31$ & $0.14\pm0.19$\\                                                     
M31N 2007-10a & \dots & \dots & \dots & \dots\\
\enddata
\end{deluxetable}

\begin{deluxetable}{crccc}
\tabletypesize{\scriptsize}
\tablenum{3}
\tablewidth{0pt}
\tablecolumns{5}
\tablecaption{Photometric Observations of M31N 2006-09c\label{phot1}}
\tablehead{\colhead{$\mathrm{HJD}$} & \colhead{} & \colhead{} & \colhead{} & \colhead{} \\ \colhead{($2,450,000+$)} & \colhead{Mag} & \colhead{Filter} & \colhead{Observer(s)} & \colhead{Telescope}}
\startdata
3993.376 &$>19.1$       &R& KH &Lelekovice  0.35-m\\
3996.404 & $18.1\pm0.2$ &R& P. Ku\v{s}nir\'{a}k& Ond\v{r}ejov  0.65-m\\
3999.609 & $17.15\pm0.1$&R& P. Ku\v{s}nir\'{a}k& Ond\v{r}ejov  0.65-m\\
4000.317 & $17.0\pm0.1$ &R& KH &Lelekovice  0.35-m\\
4000.590 & $17.35\pm0.1$&R& M. Wolf    &Ond\v{r}ejov  0.65-m\\
4001.360 & $17.3\pm0.1$ &R& KH &Lelekovice  0.35-m\\
4002.302 & $17.4\pm0.1$ &R& KH &Lelekovice  0.35-m\\
4002.329 & $17.5\pm0.1$ &R& KH &Lelekovice  0.35-m\\
4005.312 & $17.8\pm0.1$ &R& KH &Lelekovice  0.35-m\\
4007.312 & $17.9\pm0.1$ &R& KH &Lelekovice  0.35-m\\
4014.295 & $19.0\pm0.2$ &R& KH &Lelekovice  0.35-m\\
4017.258 & $19.1\pm0.2$ &R& KH &Lelekovice  0.35-m\\
4019.319 & $18.8\pm0.2$ &R& KH &Lelekovice  0.35-m\\
4024.383 & $19.3\pm0.2$ &R& KH &Lelekovice  0.35-m\\
4026.330 & $19.3\pm0.2$ &R& KH &Lelekovice  0.35-m\\
4026.364 & $19.3\pm0.2$ &R& KH &Lelekovice  0.35-m\\
4034.312 & $19.9\pm0.3$ &R& KH &Lelekovice  0.35-m\\
\\
4256.677 & $>21.2\pm0.2$ &B& MFB, MJD, AWS, KAM & LT 2.0-m \\
4260.700 & $>21.7\pm0.3$ &B& MFB, MJD, AWS, KAM & LT 2.0-m \\
\\
4260.705 & $>21.0\pm0.3$ &V& MFB, MJD, AWS, KAM & LT 2.0-m \\
\\
4260.690 & $>21.4\pm0.2$ &r& MFB, MJD, AWS, KAM & LT 2.0-m \\
\\
4260.695 & $21.8\pm0.5$  &i& MFB, MJD, AWS, KAM & LT 2.0-m \\
\enddata
\end{deluxetable}

\begin{deluxetable}{crccc}
\tabletypesize{\scriptsize}
\tablenum{4}
\tablewidth{0pt}
\tablecolumns{5}
\tablecaption{Photometric Observations of M31N 2006-10a\label{phot2}}
\tablehead{\colhead{$\mathrm{HJD}$} & \colhead{} & \colhead{} & \colhead{} & \colhead{} \\ \colhead{($2,450,000+$)} & \colhead{Mag} & \colhead{Filter} & \colhead{Observer(s)} & \colhead{Telescope}}
\startdata
4019.319 &$>19.7$       & R& KH &Lelekovice  0.35-m\\
4024.383 &$>19.0$       & R& KH &Lelekovice  0.35-m\\
4026.330 &$>20.1$       & R& KH &Lelekovice  0.35-m\\
4031.251 & $19.2\pm0.25$& R& KH &Lelekovice  0.35-m\\
4034.312 & $18.7\pm0.15$& R& KH &Lelekovice  0.35-m\\
4034.470 & $18.6\pm0.15$& R& P. Ku\v{s}nir\'{a}k& Ond\v{r}ejov  0.65-m \\
4035.360 & $18.4\pm0.3$ & R& KH &Lelekovice  0.35-m\\
4043.331 & $17.9\pm0.1$ & R& KH &Lelekovice  0.35-m\\
4047.288 & $19.0\pm0.3$ & R& KH &Lelekovice  0.35-m\\
4048.324 & $19.3\pm0.2$ & R& KH &Lelekovice  0.35-m\\
4055.296 & $18.2\pm0.15$& R& KH &Lelekovice  0.35-m\\
4055.262 & $18.2\pm0.1$ & R& P. Caga\v{s}   &Zl\'{i}n        0.28-m\\
4070.308 & $19.2\pm0.35$& R& M. Wolf, P. Zasche &Ond\v{r}ejov  0.65-m\\
4071.385 & $18.9\pm0.2$ & R& P. Ku\v{s}nir\'{a}k& Ond\v{r}ejov  0.65-m\\
4078.308 & $18.4\pm0.3$ & R& KH &Lelekovice  0.35-m\\
4078.343 & $18.6\pm0.2$ & R& KH &Lelekovice  0.35-m\\
4080.306 & $18.8\pm0.25$& R& KH &Lelekovice  0.35-m\\
4084.212 & $18.9\pm0.2$ & R& KH &Lelekovice  0.35-m\\
4093.174 & 19.5:      & R& KH &Lelekovice  0.35-m\\
4096.325 & $19.7\pm0.25$& R& KH &Lelekovice  0.35-m\\
4097.222 & $19.5\pm0.25$& R& KH &Lelekovice  0.35-m\\
4115.194 &$>20.0$       & R& KH &Lelekovice  0.35-m\\
4121.381 &$>19.8$       & R& KH &Lelekovice  0.35-m\\
4122.331 &$>19.5$       & R& KH &Lelekovice  0.35-m\\
4122.377 &$>19.4$       & R& KH &Lelekovice  0.35-m\\
\\
4044.337 & $18.07\pm0.04$ &B& MFB, MJD, AWS, KAM & LT 2.0-m\\
4050.589 & $19.06\pm0.03$ &B& MFB, MJD, AWS, KAM & LT 2.0-m\\
4056.585 & $18.88\pm0.03$ &B& MFB, MJD, AWS, KAM & LT 2.0-m\\
4062.606 & $18.58\pm0.03$ &B& MFB, MJD, AWS, KAM & LT 2.0-m\\
4069.484 & $19.24\pm0.04$ &B& MFB, MJD, AWS, KAM & LT 2.0-m\\
4071.549 & $19.06\pm0.09$ &B& MFB, MJD, AWS, KAM & LT 2.0-m\\
4074.575 & $19.20\pm0.11$ &B& MFB, MJD, AWS, KAM & LT 2.0-m\\
4077.542 & $19.26\pm0.13$ &B& MFB, MJD, AWS, KAM & LT 2.0-m\\
4084.460 & $19.24\pm0.03$ &B& MFB, MJD, AWS, KAM & LT 2.0-m\\
4092.454 & $19.93\pm0.05$ &B& MFB, MJD, AWS, KAM & LT 2.0-m\\
4099.407 & $20.03\pm0.11$ &B& MFB, MJD, AWS, KAM & LT 2.0-m\\
4101.393 & $19.78\pm0.07$ &B& MFB, MJD, AWS, KAM & LT 2.0-m\\
4114.372 & $>21.90\pm0.24$ &B& MFB, MJD, AWS, KAM & LT 2.0-m\\
4120.432 & $>20.89\pm0.28$ &B& MFB, MJD, AWS, KAM & LT 2.0-m\\
4254.685 & $>17.6\pm0.9$     &B& MFB, MJD, AWS, KAM & LT 2.0-m\\
\\
4044.334 &$ 17.90\pm0.03$ &V& MFB, MJD, AWS, KAM & LT 2.0-m\\
4050.586 &$ 19.10\pm0.02$ &V& MFB, MJD, AWS, KAM & LT 2.0-m\\
4056.582 &$ 18.80\pm0.03$ &V& MFB, MJD, AWS, KAM & LT 2.0-m\\
4062.604 &$ 18.48\pm0.02$ &V& MFB, MJD, AWS, KAM & LT 2.0-m\\
4069.481 &$ 19.25\pm0.04$ &V& MFB, MJD, AWS, KAM & LT 2.0-m\\
4071.546 &$ 19.02\pm0.12$ &V& MFB, MJD, AWS, KAM & LT 2.0-m\\
4074.572 &$ 19.39\pm0.10$ &V& MFB, MJD, AWS, KAM & LT 2.0-m\\
4077.539 &$ 19.30\pm0.05$ &V& MFB, MJD, AWS, KAM & LT 2.0-m\\
4084.457 &$ 19.16\pm0.03$ &V& MFB, MJD, AWS, KAM & LT 2.0-m\\
4092.451 &$ 19.67\pm0.05$ &V& MFB, MJD, AWS, KAM & LT 2.0-m\\
4099.404 &$ 19.64\pm0.10$ &V& MFB, MJD, AWS, KAM & LT 2.0-m\\
4101.391 &$ 19.73\pm0.06$ &V& MFB, MJD, AWS, KAM & LT 2.0-m\\
4114.370 &$ >21.55\pm0.22$ &V& MFB, MJD, AWS, KAM & LT 2.0-m\\
4120.429 &$ >21.22\pm0.12$ &V& MFB, MJD, AWS, KAM & LT 2.0-m\\
4254.692 &$ >18.4\pm0.4$ &V& MFB, MJD, AWS, KAM & LT 2.0-m\\
\\
4254.671 &$>19.0\pm0.3$ &r& MFB, MJD, AWS, KAM & LT 2.0-m\\
\enddata
\end{deluxetable}

\begin{deluxetable}{crccc}
\tabletypesize{\scriptsize}
\tablenum{5}
\tablewidth{0pt}
\tablecolumns{5}
\tablecaption{Photometric Observations M31N 2006-10b\label{phot3}}
\tablehead{\colhead{$\mathrm{HJD}$} & \colhead{} & \colhead{} & \colhead{} & \colhead{} \\ \colhead{($2,450,000+$)} & \colhead{Mag} & \colhead{Filter} & \colhead{Observer(s)} & \colhead{Telescope}}
\startdata
4043.888 &$18.95\pm0.06$ &B& MFB, MJD, AWS, KAM & LT 2.0-m\\
4049.001 &$19.55\pm0.04$ &B& MFB, MJD, AWS, KAM & LT 2.0-m\\
4057.034 &$20.39\pm0.06$ &B& MFB, MJD, AWS, KAM & LT 2.0-m\\
4062.963 &$20.89\pm0.07$ &B& MFB, MJD, AWS, KAM & LT 2.0-m\\
4068.945 &$20.99\pm0.13$ &B& MFB, MJD, AWS, KAM & LT 2.0-m\\
4071.940 &$>20.14\pm0.21$ &B& MFB, MJD, AWS, KAM & LT 2.0-m\\
4074.973 &$>20.35\pm0.20$ &B& MFB, MJD, AWS, KAM & LT 2.0-m\\
4083.911 &$21.41\pm0.10$  &B& MFB, MJD, AWS, KAM & LT 2.0-m\\
4098.938 &$>20.65\pm0.27$ &B& MFB, MJD, AWS, KAM & LT 2.0-m\\
4105.915 &$>19.74\pm0.54$ &B& MFB, MJD, AWS, KAM & LT 2.0-m\\
4113.883 &$>22.91\pm0.69$ &B& MFB, MJD, AWS, KAM & LT 2.0-m\\
4119.958 &$>21.83\pm0.38$ &B& MFB, MJD, AWS, KAM & LT 2.0-m\\
4248.204 &$>22.42\pm0.43$ &B& MFB, MJD, AWS, KAM & LT 2.0-m\\
\\
4048.998 &$20.02\pm0.05$  &V& MFB, MJD, AWS, KAM & LT 2.0-m\\
4053.817 &$>17.33\pm0.29$ &V& MFB, MJD, AWS, KAM & LT 2.0-m\\
4057.031 &$20.88\pm0.07$  &V& MFB, MJD, AWS, KAM & LT 2.0-m\\
4062.960 &$21.27\pm0.08$  &V& MFB, MJD, AWS, KAM & LT 2.0-m\\
4068.942 &$>21.69\pm0.23$ &V& MFB, MJD, AWS, KAM & LT 2.0-m\\
4071.937 &$>20.12\pm0.21$ &V& MFB, MJD, AWS, KAM & LT 2.0-m\\
4074.970 &$>20.64\pm0.19$ &V& MFB, MJD, AWS, KAM & LT 2.0-m\\
4083.908 &$>22.30\pm0.22$ &V& MFB, MJD, AWS, KAM & LT 2.0-m\\
4098.936 &$>20.51\pm0.22$ &V& MFB, MJD, AWS, KAM & LT 2.0-m\\
4101.885 &$>20.13\pm0.33$ &V& MFB, MJD, AWS, KAM & LT 2.0-m\\
4105.912 &$>17.61\pm0.60$ &V& MFB, MJD, AWS, KAM & LT 2.0-m\\
4107.986 &$>18.40\pm0.28$ &V& MFB, MJD, AWS, KAM & LT 2.0-m\\
4113.880 &$>21.96\pm0.39$ &V& MFB, MJD, AWS, KAM & LT 2.0-m\\
4119.955 &$>22.76\pm0.65$ &V& MFB, MJD, AWS, KAM & LT 2.0-m\\
4248.209 &$>22.30\pm0.43$ &V& MFB, MJD, AWS, KAM & LT 2.0-m\\
\\
4248.193 &$>22.9\pm0.4$ &r& MFB, MJD, AWS, KAM & LT 2.0-m\\
\\
4248.199 &$>21.5\pm0.3$ &i& MFB, MJD, AWS, KAM & LT 2.0-m\\
\enddata
\end{deluxetable}

\begin{deluxetable}{crccc}
\tabletypesize{\scriptsize}
\tablenum{6}
\tablewidth{0pt}
\tablecolumns{5}
\tablecaption{Photometric Observations M31N 2006-11a\label{phot4}}
\tablehead{\colhead{$\mathrm{HJD}$} & \colhead{} & \colhead{} & \colhead{} & \colhead{} \\ \colhead{($2,450,000+$)} & \colhead{Mag} & \colhead{Filter} & \colhead{Observer(s)} & \colhead{Telescope}}
\startdata
4048.324 &$>19.8$       & R& KH &Lelekovice  0.35-m\\
4055.296 &$>19.8$       & R& KH &Lelekovice  0.35-m\\
4070.263 & $16.9\pm0.1$ & R& M. Wolf, P. Zasche &Ond\v{r}ejov  0.65-m\\
4070.308 & $16.6\pm0.15$& R& M. Wolf, P. Zasche &Ond\v{r}ejov  0.65-m\\
4078.308 & $16.1\pm0.1$ & R& KH &Lelekovice  0.35-m\\
4078.343 & $16.0\pm0.1$ & R& KH &Lelekovice  0.35-m\\
4080.306 & $16.3\pm0.1$ & R& KH &Lelekovice  0.35-m\\
4084.212 & $16.9\pm0.1$ & R& KH &Lelekovice  0.35-m\\
4093.174 & $17.7\pm0.15$& R& KH &Lelekovice  0.35-m\\
4096.325 & $17.8\pm0.15$& R& KH &Lelekovice  0.35-m\\
4097.222 & $17.8\pm0.15$& R& KH &Lelekovice  0.35-m\\
4115.194 & $18.8\pm0.2$ & R& KH &Lelekovice  0.35-m\\
4121.381 & $18.9\pm0.2$ & R& KH &Lelekovice  0.35-m\\
4122.331 & $19.2\pm0.25$& R& KH &Lelekovice  0.35-m\\
4122.377 & $19.1\pm0.3$ & R& KH &Lelekovice  0.35-m\\
4126.289 & $19.3\pm0.3$ & R& KH &Lelekovice  0.35-m\\
4126.339 & $19.1\pm0.3$ & R& KH &Lelekovice  0.35-m\\
4128.275 & $19.2\pm0.25$& R& KH &Lelekovice  0.35-m\\
4135.298 & $19.4\pm0.3$ & R& KH &Lelekovice  0.35-m\\
4135.334 & $19.3\pm0.25$& R& KH &Lelekovice  0.35-m\\
4141.362 & $19.2\pm0.3$ & R& KH &Lelekovice  0.35-m\\
4146.295 & $19.4\pm0.25$& R& KH &Lelekovice  0.35-m\\
4149.273 & $19.4\pm0.25$& R& KH &Lelekovice  0.35-m\\
4166.247 & $19.7\pm0.2$ & R& KH &Ond\v{r}ejov  0.65-m\\
4167.366 & $20.0\pm0.35$& R& KH &Ond\v{r}ejov  0.65-m\\
4170.264 & $20.0\pm0.3$ & R& KH &Ond\v{r}ejov  0.65-m\\
4173.269 & $19.8\pm0.25$& R& P. Ku\v{s}nir\'{a}k, T. Henych &Ond\v{r}ejov  0.65-m\\
4174.268 &$>19.7$       & R& KH &Lelekovice  0.35-m\\
4240.561 &$>20.0$       & R& KH &Ond\v{r}ejov  0.65-m\\
\\
4140.868 &$20.12\pm0.05$ &B& MFB, MJD, AWS, KAM & LT 2.0-m\\
\\
4140.872 &$20.54\pm0.06$ &V& MFB, MJD, AWS, KAM & LT 2.0-m\\
\\
4140.856 &$19.22\pm0.03$ &r& MFB, MJD, AWS, KAM & LT 2.0-m\\
\\
4140.862 &$20.22\pm0.08$ &i& MFB, MJD, AWS, KAM & LT 2.0-m\\
\enddata
\end{deluxetable}

\begin{deluxetable}{crccc}
\tabletypesize{\scriptsize}
\tablenum{7}
\tablewidth{0pt}
\tablecolumns{5}
\tablecaption{Photometric Observations of M31N 2007-08d\label{phot5}}
\tablehead{\colhead{$\mathrm{HJD}$} & \colhead{} & \colhead{} & \colhead{} & \colhead{} \\ \colhead{($2,450,000+$)} & \colhead{Mag} & \colhead{Filter} & \colhead{Observer(s)} & \colhead{Telescope}}
\startdata
4380.401 & $18.6\pm0.2$ &R& KH, P. Ku\v{s}nir\'{a}k &Ond\v{r}ejov  0.65-m\\
4380.424 & $18.5\pm0.15$&R& KH, P. Ku\v{s}nir\'{a}k &Ond\v{r}ejov  0.65-m\\
4382.235 & $19.0\pm0.25$&R& KH, P. Ku\v{s}nir\'{a}k &Ond\v{r}ejov  0.65-m\\
4387.219 &$>19.5$       &R& KH, P. Ku\v{s}nir\'{a}k &Ond\v{r}ejov  0.65-m\\
4387.231 & $19.5\pm0.3$ &R& KH, P. Ku\v{s}nir\'{a}k &Ond\v{r}ejov  0.65-m\\
4387.561 & $19.6\pm0.2$ &R& KH &Ond\v{r}ejov  0.65-m\\
4388.227 & $19.8\pm0.25$&R& KH, P. Ku\v{s}nir\'{a}k &Ond\v{r}ejov  0.65-m\\
4388.626 & $19.6\pm0.35$&R& KH &Ond\v{r}ejov  0.65-m\\
4389.233 & $19.8\pm0.3$ &R& KH, P. Ku\v{s}nir\'{a}k &Ond\v{r}ejov  0.65-m\\
4389.646 & $20.0\pm0.3$ &R& KH &Ond\v{r}ejov  0.65-m\\
\enddata
\end{deluxetable}

\begin{deluxetable}{lcccccc}
\tabletypesize{\scriptsize}
\tablenum{8}
\tablewidth{0pt}
\tablecolumns{7}
\tablecaption{Summary of HET Spectroscopic Observations\label{hettab}}
\tablehead{\colhead{} & \colhead{R.A.} & \colhead{Decl.} & \colhead{}&\colhead{Exp.}
 & \colhead{Coverage} & \colhead{}  \\
\colhead{Nova} & \colhead{(2000.0)} & \colhead{(2000.0)} & \colhead{UT Date} & \colhead{(sec)} & \colhead{(\AA)} & \colhead{Weather}}
\startdata
M31N 2006-09c  &  00 42 42.38 & 41 08 45.5 & 24.18 Sep 2006 &1800 &4275--7250 & Spec \\
M31N 2006-10a  &  00 41 43.23 & 41 11 45.9 & 30.31 Oct 2006 &1500 &4275--7250 & Spec \\
M31N 2006-10b  &  00 39 27.38 & 40 51 09.8 & 23.24 Nov 2006 &1200 &4275--7250 & Phot\\
M31N 2006-11a  &  00 42 56.81 & 41 06 18.4 & 28.23 Nov 2006 &1200 &4275--7250 & Spec\\
M31N 2007-08d  &  00 39 30.27 & 40 29 14.2 & 13.44 Sep 2007 &1200 &4275--7250 & Spec\\
M31N 2007-10a  &  00 42 55.95 & 41 03 22.0 & 19.13 Oct 2007 &1400 &4275--7250 & Phot\\
M31N 2007-11d  &  00 44 54.60 & 41 37 40.0 & 04.22 Dec 2007 &1200 &4100--8900 & Phot\\
M31N 2007-11e  &  00 45 47.74 & 42 02 03.5 & 05.23 Dec 2007 &1200 &4100--8900 & Phot\\
\enddata
\end{deluxetable}

\begin{deluxetable}{lrrrr}
\tabletypesize{\scriptsize}
\tablenum{9}
\tablewidth{0pt}
\tablecolumns{5}
\tablecaption{Balmer Emission Line Properties\label{emission}}
\tablehead{
\colhead{} & \multicolumn{2}{c}{EW (\AA)} & \multicolumn{2}{c}{FWHM (km/s)} \\
\colhead{Nova} & \colhead{H$\beta$} & \colhead{H$\alpha$} & \colhead{H$\beta$} & \colhead{H$\alpha$}}
\startdata
M31N 2006-09c  & $-127$ & $-470$ & 1910 & 1920 \\
M31N 2006-10a  & $-44$  & $-90$  & 950  & 810 \\
M31N 2006-10b  & $-133$ & $-2130$& 3030 & 3562 \\
M31N 2006-11a  & $-35$  & $-57$  & 1420 & 1120 \\
M31N 2007-07f\tablenotemark{a} & $-56$  & $-195$ & 1340 & 1110 \\
M31N 2007-08a  & \dots  & $-66$& \dots  & 1880\\
M31N 2007-08d  & $-68 $ & $-284$ & 1160 & 1180 \\
M31N 2007-10a  & $-37 $ & $-154$ &  470 & 500 \\
M31N 2007-11d  & $-297$ & $-1223$& 2060 & 2260 \\
M31N 2007-11e  & $-137$ & $-306$ & 1750 & 1600 \\
\enddata
\tablenotetext{a}{Measurements are from a spectrum kindly provided by R. Quimby
(private communication).}
\end{deluxetable}

\begin{deluxetable}{lrrrll}
\tabletypesize{\scriptsize}
\tablenum{10}
\tablewidth{0pt}
\tablecolumns{6}
\tablecaption{Spatial Positions and Spectroscopic Types\label{spatpos}}
\tablehead{\colhead{} & \colhead{$\Delta\alpha~cos\delta$} & \colhead{$\Delta\delta$} & \colhead{$a$}& \colhead{} & \colhead{}\\
\colhead{Nova} & \colhead{($'$)} & \colhead{($'$)} & \colhead{($'$)} & \colhead{Type} & \colhead{References\tablenotemark{a}}}
\startdata
M31N 2006-09c  & $-0.37$& $-7.39$&  11.61  & Fe II & 1,2 \\
M31N 2006-10a  &$-11.50$& $-4.36$&  17.31  & Fe II & 1 \\
M31N 2006-10b  &$-37.25$&$-24.81$&  63.14  & He/N  & 1 \\
M31N 2006-11a  &  2.35 & $-9.84$&  20.46  & Fe II &  1 \\
M31N 2007-07f  & $-45.77$&$-22.95$&  81.74 & Fe II & 3\\
M31N 2007-08a  &$-20.78$&$-22.25$&  30.49  & Fe II?\tablenotemark{b}  & 4 \\
M31N 2007-08d  &$-36.91$&$-46.74$&  59.56  & Fe II & 1 \\
M31N 2007-10a  &  2.19 &$-12.78$&  27.57  & Fe II?\tablenotemark{c}  & 1, 5 \\
M31N 2007-11d  & 24.34 &  21.60 &  34.03  & Fe II & 1, 6, 7 \\
M31N 2007-11e\tablenotemark{d}  & 34.06 &  46.07 &  57.52  & Fe II & 1, 8 \\
\enddata
\tablenotetext{a}{References: (1) This work; (2) \citet{sha06}; (3) \citet{qua07}; (4) \citet{bar07}; (5) \citet{gal07}; (6) \citet{qub07}; (7) \citet{sha09}; (8) \citet{dim07}}
\tablenotetext{b}{possible Hybrid nova}
\tablenotetext{c}{possible peculiar He/N nova with narrow He I lines}
\tablenotetext{d}{Neon nova}
\end{deluxetable}

\begin{deluxetable}{lccccr}
\tabletypesize{\scriptsize}
\tablenum{11}
\tablewidth{0pt}
\tablecolumns{6}
\tablecaption{Eruption Properties and Dust Formation Timescales\label{dust}}
\tablehead{
\colhead{} & \colhead{$t_2$\tablenotemark{a}} & \colhead{$t_{\rm cond}$} & \colhead{$t_{\rm IRmax}$} & \colhead{$\Delta T$} & \colhead{}\\
\colhead{Nova} & \colhead{(days)} & \colhead{(days)} &
\colhead{(days)} & \colhead{(days)} & \colhead{Type}}
\startdata
V1370 Aql    & 15  &  $<16$ & 37 & \dots & He/N\\
V1419 Aql    & 25  &  $\sim$90  & 40 & \dots & Fe II\\
OS And       & 11  &  22 & \dots & \dots & Fe II\\ 
T Aur        & 80  &  87 & \dots & \dots & Fe II\\
V705 Cas     & 33  &  63 & 100 & \dots & Fe II\\
V842 Cen     & 43  &  36 & 87 & \dots & Fe II\\
V476 Cyg     & 7   &  40 & \dots & \dots & Fe II\\
V1668 Cyg    & 11  &  33 & 57 & \dots & Fe II\\
V2274 Cyg    & 22  &  40 & \dots & \dots & Fe II\\
DQ Her       & 76  &  105 & \dots & \dots & Fe II\\
V827 Her     & 21  &  43 & \dots & \dots &  \dots\\
V838 Her     & 1.2 &  18 & 25 & \dots & He/N\\
V445 Pup     & 215 &  218 & \dots & \dots & \dots\\
V732 Sgr     & 65  &  73 & \dots & \dots & \dots\\
FH Ser       & 49  &  60 & 90 & \dots & Fe II\\
LW Ser       & 32  &  23 & 75 & \dots & Fe II\\
V992 Sco     & 100 &  93 & \dots & \dots & Fe II\\
NQ Vul       & 21  &  62 & 80 & \dots & Fe II\\
PW Vul       & 44  &  154 & 280 & \dots & Fe II\\
QU Vul       & 20  &  40  & 240 & \dots & Fe II\\
QV Vul       & 37  &  56 & 115 & \dots & Fe II\\
LMC 1998\#1  & 23  &  59 & 57 & \dots &\dots\\
M31N 2006-09c &  $17(R)$  & \dots & \dots & 154.61 & Fe II\\
M31N 2006-10a &  $74(B),83(V),112(R)$  &  $<116$ & \dots & 116 & Fe II\\
M31N 2006-10b &  $20(B),6(R)$  & \dots & \dots & 110.23 & He/N\\
M31N 2006-11a &  $21(R)$  &  \dots & \dots & 86.25 & Fe II\\
M31N 2007-07f &  $50:(R)$ & $<203$ & \dots & 203 & Fe II\\
M31N 2007-08a &  2        & \dots & \dots & 158.48 & Fe II?\\
M31N 2007-08d &  $14(R)$  & \dots & \dots & 179.9 & Fe II\\
M31N 2007-10a &  7        & \dots & \dots & 119.62 & Fe II?\\
M31N 2007-11d &  $14(B),11(V),10(R)$  & \dots & \dots & 98.37 & Fe II\\
M31N 2007-11e &  $27:(R)$ & \dots & \dots & 86.8 & Fe II\\
\enddata
\tablenotetext{a}{Galactic nova $t_2$ times from \citet{str10}}
\end{deluxetable}

%% Tables may also be prepared as separate files. See the accompanying
%% sample file table.tex for an example of an external table file.
%% To include an external file in your main document, use the \input
%% command. Uncomment the line below to include table.tex in this
%% sample file. (Note that you will need to comment out the \documentclass,
%% \begin{document}, and \end{document} commands from table.tex if you want
%% to include it in this document.)

%% \input{table}

%% The following command ends your manuscript. LaTeX will ignore any text
%% that appears after it.

\end{document}